\documentclass[a4paper, reprint, aps, nofootinbib, groupedaddress]{revtex4-2} % arXiv

\usepackage{graphicx}
\usepackage{physics,mathrsfs,mathtools,array,bm}
\usepackage{amsmath,amsthm,amssymb,amsfonts}
\usepackage{newtxtext,newtxmath}
\usepackage{orcidlink}
\usepackage{hyperref}
\hypersetup{
  pdftitle ={Shape dependence of Edelstein and magnetoelectric effects in the V-shaped model}, 
  pdfauthor={Shuhei Kanda, Satoru Hayami}, 
  colorlinks=true, linkcolor=blue, urlcolor=blue, citecolor=blue
}

\begin{document}

%- title -----------------------------------------------------------------------------------------------------------------

\title{Shape dependence of Edelstein and magnetoelectric effects in the V-shaped model}
\author{Shuhei Kanda\,\orcidlink{0009-0007-0852-9979}}\email[Contact author: ]{kanda@phys.sci.hokudai.ac.jp} 
% \author{Shuhei Kanda}
\author{Satoru Hayami\,\orcidlink{0000-0001-9186-6958}} 
% \author{Satoru Hayami}
\affiliation{Graduate School of Science, \href{https://ror.org/02e16g702}{Hokkaido University}, Sapporo 060-0810, Japan \vspace{8pt}}
% \affiliation{Graduate School of Science, Hokkaido University, Sapporo 060-0810, Japan}
\date{\today}

\begin{abstract}
We theoretically investigate the shape dependence and microscopic mechanism of the magnetoelectric (ME) effect, including both nonmagnetic (Edelstein-type) and magnetic origins, in a V-shaped one-dimensional chain model.
Numerical calculations based on the Kubo formula reveal that the nonmagnetic-driven ME response is maximized at an apex angle of $\theta \approx 0.6\pi$.
To clarify its origin, we derive a low-energy effective Hamiltonian in the $s$-orbital subspace and demonstrate that the polarity induced by the V-shaped geometry manifests as an effective spin--orbit interaction.
An analytical derivation of the Green's function shows that the geometric effect can be described as a $T$-matrix contribution associated with local symmetry breaking.
This scattering-theoretical representation enables us to identify the contributions of local geometric symmetry breaking to the ME response.
Using a multipole-basis representation, we identify symmetry-based selection rules for the ME tensor and show that the coupling between the effective spin--orbit interaction and the orbital angular momentum generated across the apex plays an essential role.
The resulting angular dependence, $\sin{\theta}\sin{\theta/2}$, peaks at $\theta = 2\tan^{-1}\sqrt{2} \approx 0.608\pi$, in good agreement with the numerical results.
We also analyze a ferromagnetic V-shaped model including the Zeeman interaction and show that the magnetic-driven ME response originates from the spin magnetization induced by the coupling between the electric-field--driven charge-potential gradient and the Zeeman term.
These results reveal distinct ME mechanisms depending on the presence or absence of time-reversal symmetry and clarify the microscopic origin of shape-dependent ME responses in V-shaped chains.
\end{abstract}

\maketitle

%- main-  -----------------------------------------------------------------------------------------------------------------
\section{introduction} \label{sec:introduction} %--------------------------------------------------------------------------

Artificially structured metamaterials (or metasurfaces) have enabled unprecedented creation and manipulation of physical phenomena~\cite{cai2007optical, soukoulis2010optical, soukoulis2011past, zheludev2012metamaterials, https://doi.org/10.1002/smtd.201600064}.
In particular, recent advances in microfabrication techniques have made it possible to induce specific symmetries into a system, for example, by fabricating holes of designed shapes~\cite{matsubara2022polarization, PhysRevB.107.155419, Cavannaetal2025} or
by constructing tailored nanostructured architectures~\cite{PhysRevLett.95.227401, soukoulis2010optical, yu2011light, gartside2021reconfigurable,10.1063/5.0089235}.
Similarly, in atomic-scale and nanoscale materials, the control of geometric configurations and atomic arrangements provides opportunities to tailor electronic states and physical properties through structural and symmetry engineering~\cite{Link1999,Jain2006,cai2010atomically,cao2018correlated,Houtsma2021}.
Such artificial systems provide a valuable platform for exploring and realizing exotic physical phenomena, which can be flexibly engineered through the geometric parameters of the constituent nanostructures, including their composition, shape, periodicity, and mutual arrangement.

To systematically characterize the symmetry and associated physical responses of such artificially engineered systems, it is useful to adopt a symmetry-based theoretical framework.
The finite multipole moments in a system reflect its underlying symmetry.
The multipole description provides a powerful way to connect macroscopic phenomena, microscopic degrees of freedom, and symmetry principles~\cite{doi:10.7566/JPSJ.87.033709, PhysRevB.98.165110, PhysRevB.98.245129, PhysRevB.104.054412,kusunose2022generalization,doi:10.7566/JPSJ.91.014701,hayami2024unified}.
Microscopic internal degrees of freedom, such as spin, orbital, and charge, can be systematically classified within the multipole basis, 
which provides a complete set for describing an arbitrary electronic states in a given Hilbert space~\cite{doi:10.7566/JPSJ.87.033709,doi:10.7566/JPSJ.89.104704,kusunose2022generalization,hayami2024unified}.
Consequently, extensive studies have been devoted to macroscopic response phenomena originating from bulk electronic degrees of freedom using multipole basis.
These efforts have employed the multipole basis to extract key model parameters and to characterize the resulting emergent behaviors~\cite{PhysRevB.102.144441,doi:10.7566/JPSJ.91.014701,kusunose2022generalization,hayami2024unified}.

Meanwhile, the detailed correspondence between the symmetries induced by nanoscale geometries and the multipole degrees of freedom remains unclear, 
beyond the level of symmetry classification~\cite{PhysRevB.98.165110,PhysRevB.104.054412,hayami2024unified}.
For example, it is still unclear how local geometric structures affect electronic states and through what microscopic mechanisms they influence macroscopic physical properties.
Elucidating the interplay between nanoscale geometries and multipole degrees of freedom may provide valuable insights into the understanding and design of geometry-induced electronic properties in nanostructures.

We focus on a V-shaped (so-called ``ku-noji''-shaped in Japanese) structure~\cite{7109859,xie2019anomalous,hu2021review,Billot15,Babynina2016}, 
which possesses $C_{\mathrm{2v}}\,(\mathrm{m2m})$ symmetry or $C_{\mathrm{s}}\,(\mathrm{m})$ symmetry when the substrate is taken into account.
In such metamaterials, the shape-induced polarity gives rise to a finite electric dipole moment, which becomes the microscopic origin of
the Edelstein effect~\cite{EDELSTEIN1990233,furukawa2017observation, gonzalez2024non, gkt2-x7mm},
the nonlinear Hall effect~\cite{PhysRevLett.115.216806,PhysRevB.95.235434,ma2019observation,wang2019ferroelectric,doi:10.7566/JPSJ.91.014701},
the nonlinear optical response~\cite{nakamura2017shift,PhysRevB.96.241203,shin2020dynamical,okyay2022second},
and the piezoelectric effect~\cite{resta2007theory,cui2018two,10.1063/5.0251679}.  
Furthermore, in ferromagnetic V-shaped systems, 
magnetic dipole degrees of freedom originating from spontaneous magnetization become active, 
leading to a multiferroic state that simultaneously breaks time-reversal and spatial-inversion symmetries. 
In such multiferroics, magnetic toroidal dipole degrees of freedom, which are characterized by polar vectors that are odd under time-reversal symmetry~\cite{DUBOVIK1990145,Spaldin_2008,kopaev2009toroidal, doi:10.7566/JPSJ.87.033709,hayami2024unified}, play an essential role.
These toroidal moments give rise to rich physical phenomena such as 
magnetoelectric (ME) effects~\cite{DUBOVIK1990145,Fiebig_2005,Spaldin_2008, hayami2016emergent,doi:10.7566/JPSJ.87.033702,PhysRevB.97.134423,ding2021field,PhysRevB.105.155157,PhysRevB.111.L201112},
nonreciprocal conductivity~\cite{tokura2018nonreciprocal,doi:10.7566/JPSJ.91.115001,PhysRevB.105.155157,doi:10.7566/JPSJ.94.083705},
and nonreciprocal directional dichroism~\cite{PhysRevLett.95.237402,doi:10.1143/JPSJ.81.023712,PhysRevB.103.L180410,10.1063/5.0089235}.

In this paper, we investigate the shape dependence and the underlying mechanism of the ME response arising from the V-shaped geometry using a one-dimensional (1D) chain model; 
hereafter, the term ``ME effect'' is used in a generalized sense to refer collectively to the nonmagnetic-driven Edelstein effect associated with electric dipoles and the magnetic-driven ME response induced by magnetic toroidal dipoles. 
The finite V-shaped chain serves as a minimal microscopic model for examining local geometric symmetry breaking, rather than a material-specific description of a bulk crystal or a mesoscopic device.
Here, we aim to clarify how local geometric symmetry breaking is encoded in ME responses from a microscopic viewpoint.
First, we evaluate the shape dependence of the nonmagnetic-driven ME effect in a model including $s$ and $p$ orbitals. 
We find that the magnitude of the ME effect reaches its maximum at an apex angle of $\theta \approx 0.6\pi$. 
To analyze the mechanism underlying this dependence, we next perform a more detailed model analysis.
By projecting onto the $s$-orbital basis, we construct a low-energy effective Hamiltonian, in which the polarity induced by the V-shaped geometry emerges as an effective spin--orbit interaction. 
Furthermore, by analyzing the shape dependence of the Green's function, we clarify that the geometric effect manifests itself as a $T$-matrix contribution to the Green's function.
We then analyze the selection rules using the multipole basis, which reveals that the effective spin--orbit interaction and the orbital angular momentum generated by motion across the apex angle play essential roles in the ME effect. 
The angular dependence of this mechanism takes the form $\sin{\theta} \sin\theta/2$, exhibiting a peak around $\theta = 2 \tan^{-1}\sqrt{2} \approx 0.608\pi$, 
demonstrating that the observed shape dependence is a direct consequence of the underlying microscopic mechanism.
Finally, we investigate a V-shaped ferromagnetic model including Zeeman interaction.
We find that the magnetic-driven ME effect in this case arises from spin magnetization induced by the coupling between the charge-potential gradient generated by the V-shaped geometry and the Zeeman interaction. 
These distinct physical mechanisms result in different dependence of the ME effect on both shape and system size.
These results provide microscopic insights into the relationship between local geometric symmetry breaking and ME responses in V-shaped nanostructures.

The organization of this paper is as follows.
In Sec.~\ref{sec:sp}, we present the numerical results demonstrating the shape dependence of the nonmagnetic-driven ME effect in a model including $s$ and $p$ orbitals.
In Sec.~\ref{sec:seff}, we derive an effective Hamiltonian and show that the polarity induced by the V-shaped geometry results in the effective spin--orbit interaction.
In Sec.~\ref{sec:ShapeDip}, we analyze the shape dependence of the Green's function.
In Sec.~\ref{sec:Selection}, we employ the multipole basis to analyze the selection rules for the ME tensor, thereby clarifying the key physical mechanisms underlying the shape dependence.
In Sec.~\ref{sec:Ferromag}, we further investigate the shape dependence and the underlying mechanism of the magnetic-driven ME effect in a ferromagnetic system without the time-reversal symmetry.
Section~\ref{sec:Summary} is devoted to the summary of the present findings.
Appendixes include the derivation of the Green's function of the straight 1D chain in Appendix~\ref{sec:app:1DChaG}, the matrix representation of multipoles in the V-shaped model in Appendix~\ref{sec:app:MpB}, and the expression of the ME tensor based on the multipole basis in Appendix~\ref{sec:app:MpBME}.

\section{$s$-$p$ model} \label{sec:sp} %--------------------------------------------------------------------------

We evaluate the shape dependence of the ME effect in a model including $s$ and $p$ orbitals.  
In Sec.~\ref{subsec:spHam}, we introduce the $s$--$p$-orbital model in the V-shaped chain system.
We also show the expression of the linear ME tensor, which describes the correlation between the electric polarization and the magnetization in Sec.~\ref{subsec:spResults}.

\subsection{Model and Method}\label{subsec:spHam} %-  -  -  -  -  -  -  -  -  -  -  -  -  -  -  -  -  -

\begin{figure}[t]
  \centering
  \includegraphics[width=0.5\linewidth]{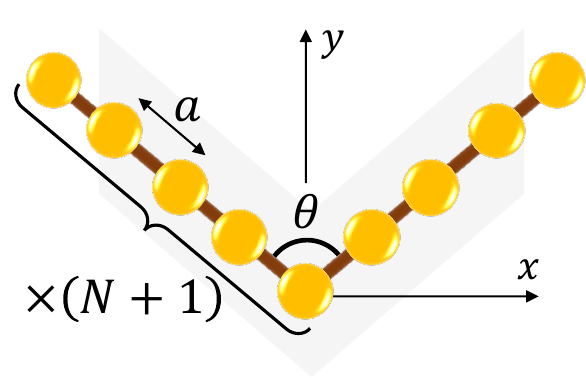} % arXiv
  \caption{\label{fig:kulat} 
  The lattice of the $2N+1$ -site $(N=4)$ V-shaped 1D chain defined in Eq.~(\ref{eq:kulat}). 
  $a$ is the lattice constant and $\theta$ is the apex angle.
  }
\end{figure}

We consider a minimal model of a V-shaped finite chain on the $xy$ plane that captures the essential symmetry and orbital structure, incorporating $s$ and $p$ orbitals, as shown in Fig.~\ref{fig:kulat}.
The tight-binding Hamiltonian is given by
\begin{align}
  H
  &=\sum_{\bm{R}_n\bm{R}_{n'}}\sum_{mm'\sigma\sigma'}
  \varphi_{m\sigma}^{\dagger}(\bm{R}_n)\,h_{mm'\sigma\sigma'}^{nn'}\,\varphi_{m'\sigma'}(\bm{R}_{n'}),
\end{align}
where $\varphi_{m\sigma}(\bm{R}_n)$ and $\varphi_{m\sigma}^{\dagger}(\bm{R}_n)$ are the fermionic annihilation and creation operators of
the $s$--$p$ orbital $m=s,p_x,p_y,p_z$,
the spin $\sigma=\uparrow,\downarrow$,
and the position $\bm{R}_n$.
The V-shaped 1D chain lattice consists of the $2N+1$ sites, whose positions are specified by $\bm{R}_n$ defined as
\begin{align}\label{eq:kulat}
  \bm{R}_n= a  \left( n \sin \frac{\theta}{2}, \abs{n} \cos\frac{\theta}{2} \right) \quad \left(-N \leq n \leq N \right) ,
\end{align}
where $a$ is the lattice constant and $\theta$ is the apex angle of the V shape, as shown in Fig.~\ref{fig:kulat}.

The Hamiltonian is decomposed into three parts as follows:
\begin{align}\label{eq:spham}
  H=H_{\mathrm{hop}}+H_{\Delta}+H_{\mathrm{SOC}} .
\end{align}
The first term $H_{\mathrm{hop}}$ represents the hopping term,
which includes the Slater-Koster parameters $t_{ss\sigma}$ and $t_{sp\sigma}$ for the nearest-neighbor hopping~\cite{Slater_PhysRev.94.1498}.
The hopping between the $p$ orbitals is neglected ($t_{pp\sigma}=t_{pp\pi}=0$) for simplicity.
$H_{\mathrm{hop}}$ is given by
\begin{align}\label{eq:hop}
  H_{\mathrm{hop}}
  &=\sum_{\bm{R},\bm{\delta}}\sum_{mm'\sigma}
  {\varepsilon}_{\bm{\delta},mm'} \, \varphi_{m\sigma}^{\dagger}(\bm{R}+a\bm{\delta}) \, \varphi_{m'\sigma}(\bm{R}),
\end{align}
where the lattice number $n$ of $\bm{R}_n$ is omitted for notational simplicity (e.g., $\bm{R}=\bm{R}_n,\bm{\delta}=\bm{\delta}_n$).
$\bm{\delta}$ is the direction cosine for the nearest-neighbor sites from $\bm{R}$.
The relationship between ${\varepsilon}$ and the Slater-Koster parameters is given by
\begin{align}\begin{aligned}
  \varepsilon_{\bm{\delta}, s s}  &= -t_{ss\sigma}, \\
  \varepsilon_{\bm{\delta}, s p_i}&= 
 -\varepsilon_{\bm{\delta}, p_i s} =  t_{sp\sigma} \delta^{i} \quad (i=x,y),\\
  \varepsilon_{\bm{\delta}, m m'} &= 0 \quad (\mathrm{otherwise}).
\end{aligned}\end{align}

The second term $H_{\Delta}$ represents the energy level of the $p$ orbital measured from that of the $s$ orbital,
which is expressed as
\begin{align}
  H_{\Delta}
  = \Delta_p \sum_{\bm{R}}\sum_{m\sigma}^{m=p_x,p_y,p_z}
  \, \varphi_{m\sigma}^{\dagger}(\bm{R}) \, \varphi_{m\sigma}(\bm{R}).
\end{align}

The third term $H_{\mathrm{SOC}}$ represents the atomic spin--orbit coupling (SOC) term,
which is represented by
\begin{align}
  H_{\mathrm{SOC}}  
  =  \frac{2\lambda}{\hbar^2} \sum_{\bm{R}} \sum_{mm'\sigma\sigma'} 
    \bm{l}_{mm'}\cdot\bm{s}_{\sigma\sigma'} \,
    \varphi_{m\sigma}^{\dagger}(\bm{R})\,\varphi_{m'\sigma'}(\bm{R}),
\end{align}
where $\bm{l}_{mm'}$ and $\bm{s}_{\sigma\sigma'}$ 
represent the matrix elements of the orbital and spin angular momentum operators, respectively.
$\hbar$ is the reduced Planck constant.
It is noted that the orbital angular momentum operator is defined within the $p$-orbital subspace.

For the above model, we evaluate the ME tensor by using the linear response theory.
The ME tensor $\alpha^{\nu;\rho}$, which corresponds to the coefficient describing the linear response of the magnetization $\delta M^{\nu}$ to an external electric field $E^{\rho}$, is defined as:
\begin{align}
  \delta M^{\nu} = \sum_{\rho} \alpha^{\nu;\rho} E^{\rho}  .
\end{align}
Within the framework of linear response theory, the ME tensor incorporates contributions from the Edelstein effect with the dissipation, which preserves time-reversal symmetry, and from the conventional ME effect without the dissipation, which requires its breaking~\cite{PhysRevB.98.165110, PhysRevB.98.245129, PhysRevB.104.054412,hayami2024unified}.
In the V-shaped model preserving time-reversal symmetry, the point-group symmetry of the system is $C_{\mathrm{2v}}$ owing to the emergence of electric polarization along the $y$ direction.
This polar symmetry allows a finite $\alpha^{z;x}$ component in two-dimensional systems. 

\begin{figure*}[t]
  \centering
  \includegraphics{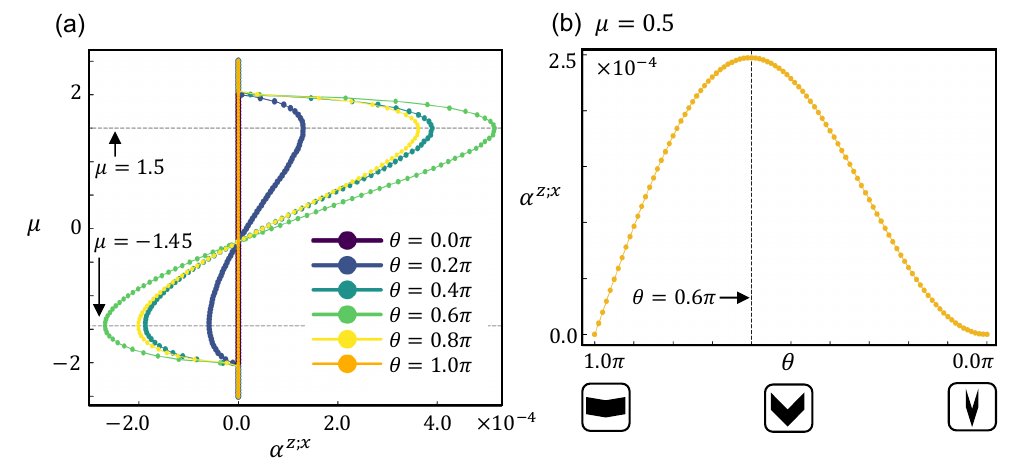}
  \caption{\label{fig:Qysp}
  (a) 
  Chemical potential $\mu$ dependence of the nonmagnetic-driven ME tensor component $\alpha^{z;x}$ for various apex angles, $\theta=0.0\pi,0.2\pi,\ldots,1.0\pi$.
  $\alpha^{z;x}$ reaches its maximum around $\mu \approx 1.5$ and its minimum around $\mu \approx -1.45$, independent of the value of $\theta$.
  (b) 
  Apex angle $\theta$ dependence of the ME tensor at $\mu=0.5$.
  The magnitude of the ME response induced by the V-shaped geometry reaches its maximum at $\theta \approx 0.6\pi$.
  In both (a) and (b), the parameters are set to $t_{ss\sigma}=1, t_{sp\sigma}=0.7, \Delta_{p}=10, \lambda=0.4, N=128, \gamma=0.05$, and $T=0.01$.
  }
\end{figure*}

We calculate the linear ME tensor using the Kubo formula:
\begin{align}\label{eq:KuboFml}
  \alpha^{\nu;\rho} = 
  -\frac{1}{V}\sum_{AB}\frac{f(E_{A})-f(E_{B})}{E_{A}-E_{B}+i\hbar\gamma}
  {M}^{\nu}_{AB} {P}^{\rho}_{BA}  ,
\end{align}
where  
$E_{A}$ are the eigenvalues of Hamiltonian matrix $\bar{h}$,
$f(E)$ is the Fermi distribution function, 
$V=(2N+1)a $ is the volume of the system,
and $\gamma$ is the scattering rate, characterized as the inverse of the relaxation time under the relaxation time approximation.
${M}^{\nu}_{AB}$ and ${P}^{\rho}_{BA}$ are matrix elements of the magnetization and polarization operators for eigenstates $A$ and $B$, respectively. 
Their operator expressions are given by
\begin{align}
  M^{\nu} 
  &= \frac{\mu_{\mathrm{B}}}{\hbar} \left( L^{\nu} + \mathrm{g} S^{\nu} \right),\\
  L^{\nu}
  &= \sum_{\bm{R}}\sum_{mm'\sigma} {l}^{\nu}_{mm'}\,\varphi_{m\sigma}^{\dagger}(\bm{R})\,\varphi_{m'\sigma}(\bm{R}),\\
  S^{\nu}
  &= \sum_{\bm{R}}\sum_{m\sigma\sigma'} {s}^{\nu}_{\sigma\sigma'}\,\varphi_{m\sigma}^{\dagger}(\bm{R})\,\varphi_{m\sigma'}(\bm{R}),\\
  P^{\rho} \label{eq:pos}
    &= e \sum_{\bm{R}} \sum_{m\sigma} 
    R^{\rho} \,
    \varphi_{m\sigma}^{\dagger}(\bm{R})\,\varphi_{m\sigma}(\bm{R}),
\end{align}
where $e$, $\mu_{\mathrm{B}}$, and $\mathrm{g}$ are the elementary charge, the Bohr magneton, and the $\mathrm{g}$ factor, respectively.
We neglect the additional contribution from 
the modern theory of orbital magnetization~\cite{PhysRevLett.102.146805,PhysRevB.83.085108} and 
interorbital polarization~\cite{gmnv-cwvr}.

\subsection{Numerical results}\label{subsec:spResults} %-  -  -  -  -  -  -  -  -  -  -  -  -  -  -  -  -  -

We evaluate the shape dependence of the ME tensor $\alpha^{z;x}$ in the V-shaped $s$--$p$ model as Eq.~(\ref{eq:spham}).
In the following numerical calculations in this subsection, we set 
\begin{align}\label{eq:HamPara}
  t_{ss\sigma}=1, t_{sp\sigma}=0.7, \Delta_{p}=10, \lambda=0.4,
\end{align}
and $N=128$.
We also set $e=\hbar=\mu_{\mathrm{B}}=a=1$, $\mathrm{g}=2$, $\gamma=0.05$, and the temperature $T=0.01$. 
Since we consider the nonmagnetic system with time-reversal symmetry, the ME response $\alpha^{z;x}$ becomes nonzero only under the finite dissipation.

First, we present the chemical potential $\mu$ dependence for various apex angles $\theta$. 
We focus on the range $-2 \leq \mu \leq 2$, in which the electronic states are primarily characterized by the $s$ orbital owing to the large energy separation $\Delta_{p}$.
As shown in Fig.~\ref{fig:Qysp}(a), $\alpha^{z;x}$ reaches its maximum around $\mu \approx 1.5$ and its minimum around $\mu \approx -1.45$, independent of the value of $\theta$. 
The disappearance of $\alpha^{z;x}$ for $\theta=\pi$ is owing to the presence of the spatial inversion symmetry, where the system reduces to the straight 1D chain. 
Meanwhile, the vanishing of $\alpha^{z;x}$ at $\theta=0$ implies that a finite opening along the $x$ direction in the V-shaped geometry, which results in the $x$-directional electric polarization, is essential for the nonmagnetic-driven ME effect.
These results also suggest that, in the present model, the characteristic shape dependence of the ME response is not restricted to a particular chemical potential.

Figure~\ref{fig:Qysp}(b) shows the $\theta$ dependence of the ME effect at $\mu=0.5$.
We choose this chemical potential as a representative value away from the extrema of the ME response. This choice is not based on any specific material or electron filling.
The magnitude of the ME response exhibits a nearly monotonic dependence on $\theta$, showing a clear peak at an apex angle of approximately $\theta \approx 0.6\pi$. 
This behavior is clarified by employing the approach based on an effective model in the subsequent section.

\section{Effective model} \label{sec:seff} %--------------------------------------------------------------------------

\begin{figure*}[t]
  \centering
  \includegraphics[width=0.95\linewidth]{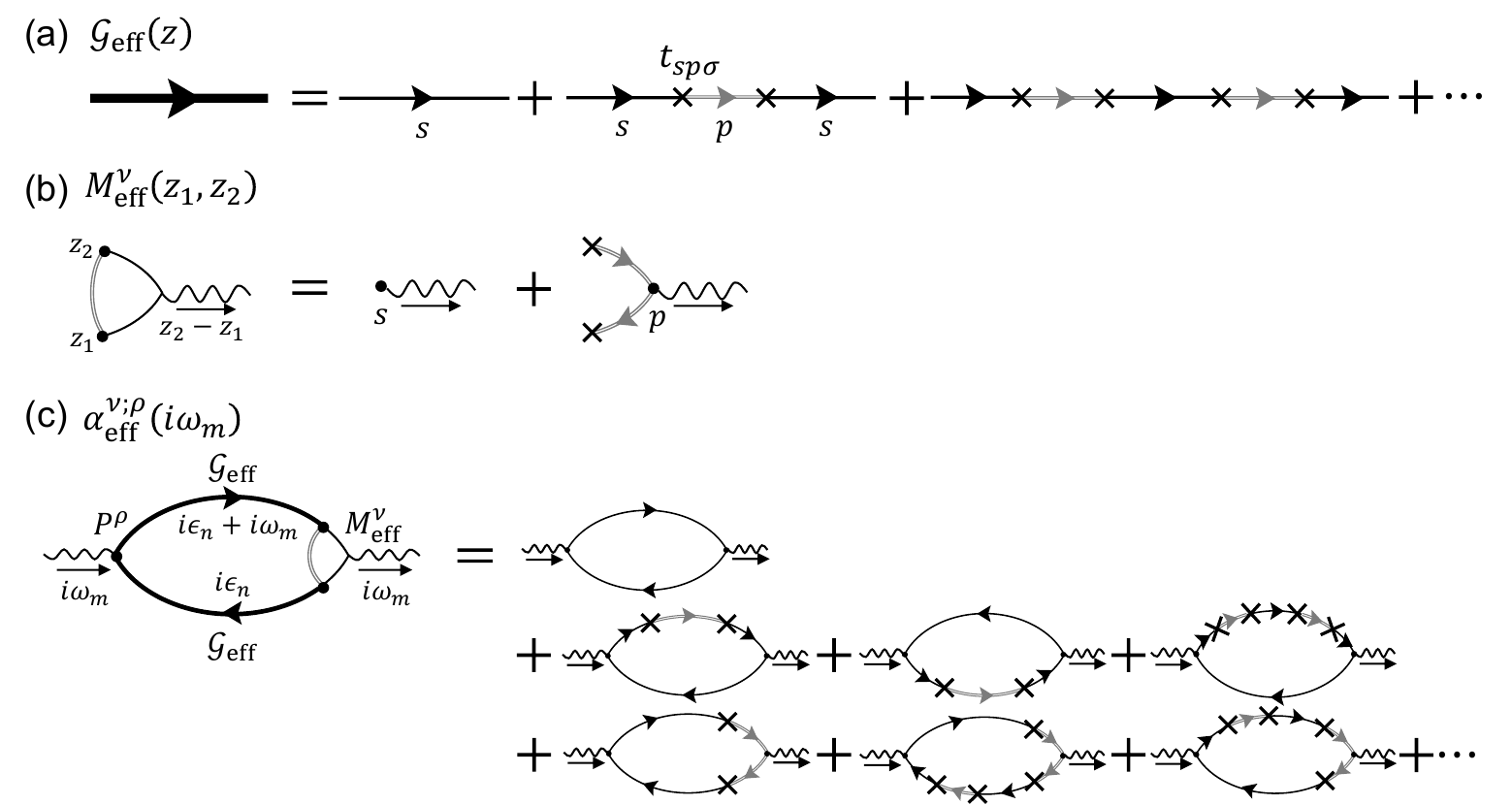}
  \caption{\label{fig:effsusFD}
  (a) 
  Diagrams representing the effective Green's function in Eq.~(\ref{eq:effgre}).
  The thick black line represents the effective Green's function $\bar{\mathcal{G}}_{\mathrm{eff}}$ in Eq.~(\ref{eq:effgre}), 
  the thin black line denotes the Green's function propagating between $s$ orbitals $\bar{g}_{\parallel}$, 
  and the gray double line represents the Green's function propagating between $p$ orbitals $\bar{g}_{\perp}$.
  The cross symbol ($\times$) indicates the $s$--$p$ orbital hybridization $\bar{\eta}$.
  As shown in Eq.~(\ref{eq:geffhyb}), the effective Green's function represents the result of incorporating the $p$-orbital effect into the $s$-orbital Green's function through the $s$--$p$ hybridization.
  (b) 
  Diagrams representing the effective magnetization operator in Eq.~(\ref{eq:effMag}).
  The vertex accompanied by a wavy line represents the magnetization operator $\bar{M}^{\nu}$.
  (c) 
  Diagrams representing the effective ME tensor in Eq.~(\ref{eq:effsus}).
  The $p$-orbital contribution is included via the $s$--$p$ hybridization in the bubble diagram closed within the $s$-orbitals.
  }
\end{figure*}

To analyze the underlying physical mechanism of the numerically obtained ME response, we adopt an effective model approach that eliminates redundant degrees of freedom to capture the essential physics.
To this end, we derive an effective Hamiltonian by projecting the $s$--$p$ orbital model onto the $s$-orbital subspace~\cite{10.1063/1.1724312,RevModPhys.36.1076,LEINAAS197819,10.1063/1.4904200}:
\begin{align}
  H_{\mathrm{eff}}(z)
  &= \sum_{\bm{R}_n\bm{R}_{n'}}\sum_{\sigma\sigma'}
  \varphi_{s\sigma}^{\dagger}(\bm{R}_n)\left[\bar{h}_{\mathrm{eff}}(z)\right]_{\sigma\sigma'}^{nn'}\varphi_{s\sigma'}(\bm{R}_{n'}),\\
  \bar{h}_{\mathrm{eff}}(z) \label{eq:effHam}
  &=  \bar{P}_{s}  \bar{h}  \bar{P}_{s}
  + \bar{\eta}^{\dagger} \, \bar{g}_{\perp}(z) \, \bar{\eta},
\end{align}
where 
$ 
  \bar{g}_{\perp}(z) = (\hbar z - \bar{Q}_{p}  \bar{h}  \bar{Q}_{p} )^{-1},\,
  \bar{\eta}= \bar{Q}_{p} \bar{h} \bar{P}_{s} ,
$
$\bar{h}$ is the matrix of the $s$--$p$ Hamiltonian in Eq.~(\ref{eq:spham}), and $ \bar{P}_{s} $ and $ \bar{Q}_{p} $ are the projection operators for $s$-orbital and $p$-orbital bases, respectively.
The effective Hamiltonian is thereby decomposed into three parts as follows:
\begin{align}\label{eq:effham}
  H_{\mathrm{eff}}(z)=H^{ss\sigma}_{\mathrm{hop}}+H^{\mathrm{2nd}}_{\mathrm{hop}}(z)+H^{\mathrm{eff}}_{\mathrm{SOI}}(z).
\end{align}
The first term $H^{ss\sigma}_{\mathrm{hop}}$ represents the nearest-neighbor hopping between $s$ orbitals in Eq.~(\ref{eq:hop}).
The second term $H^{\mathrm{2nd}}_{\mathrm{hop}}(z)$ represents the 2nd-nearest-neighbor hopping through the $s$--$p$ hybridization,
which is given by
\begin{align}
  H^{\mathrm{2nd}}_{\mathrm{hop}}(z)
  &= \sum_{\bm{R},\bm{\delta}^{\mathrm{i}}\bm{\delta}^{\mathrm{o}}}\sum_{\sigma}
  \varepsilon_{\bm{\delta}^{\mathrm{i}}\bm{\delta}^{\mathrm{o}}} \,
  \varphi_{s\sigma}^{\dagger}(\bm{R}+a \bm{\delta}^{\mathrm{o}}) \varphi_{s\sigma}(\bm{R}-a \bm{\delta}^{\mathrm{i}}),
\end{align}
where
\begin{subequations}\label{eq:2NNHdep}\begin{align}
  \varepsilon_{\bm{\delta}^{\mathrm{i}}\bm{\delta}^{\mathrm{o}}}
  &=-u(z)\,(\bm{\delta}^{\mathrm{i}}\cdot\bm{\delta}^{\mathrm{o}}), \\
  u(z) 
  &= \frac{-t_{sp\sigma}^2(\hbar z-\Delta_{p}+\lambda)}{(\hbar z-\Delta_{p}-\lambda)(\hbar z-\Delta_{p}+2\lambda)},
\end{align}\end{subequations}
and 
$\bm{\delta}^{\mathrm{i}}$ and $\bm{\delta}^{\mathrm{o}}$ are the direction cosines to the nearest-neighbor sites of $\bm{R}$.

The third term $H^{\mathrm{eff}}_{\mathrm{SOI}}(z)$ represents the effective spin--orbit interaction, which is also induced by the $s$--$p$ hybridization as
\begin{align}\begin{aligned}
  &H^{\mathrm{eff}}_{\mathrm{SOI}}(z)\\
  &= \frac{2\alpha(z)}{\hbar^2}
  \sum_{\bm{R},\bm{\delta}^{\mathrm{i}}\bm{\delta}^{\mathrm{o}}}\sum_{\sigma\sigma'}
  \bm{l}_{\bm{\delta}^{\mathrm{i}}\bm{\delta}^{\mathrm{o}}}\cdot {\bm{s}}_{\sigma\sigma'} \,
  \varphi_{s\sigma}^{\dagger}(\bm{R}+a \bm{\delta}^{\mathrm{o}}) \varphi_{s\sigma'}(\bm{R}-a \bm{\delta}^{\mathrm{i}}),
\end{aligned}\end{align}
where
\begin{subequations}\label{eq:eSOIdep}\begin{align}
  \bm{l}_{\bm{\delta}^{\mathrm{i}}\bm{\delta}^{\mathrm{o}}} 
  &= i\hbar(\bm{\delta}^{\mathrm{i}}\times\bm{\delta}^{\mathrm{o}}),\\
  \alpha(z) 
  &= \frac{t_{sp\sigma}^2\lambda}{(\hbar z-\Delta_{p}-\lambda)(\hbar z-\Delta_{p}+2\lambda)}.
\end{align}\end{subequations}
Effective spin--orbit interaction $H^{\mathrm{eff}}_{\mathrm{SOI}}$ arises from the coupling between the orbital angular momentum generated by 
two successive $sp\sigma$ hoppings and the on-site atomic spin--orbit coupling in the absence of local spatial inversion symmetry with nonzero $\bm{\delta}^{\mathrm{i}}\times\bm{\delta}^{\mathrm{o}}$.
Reflecting this, in our effective model, $\bm{l}_{\bm{\delta}^{\mathrm{i}}\bm{\delta}^{\mathrm{o}}}$ takes a finite value only at the V-shaped edge at $n=0$ within the nearest-neighbor $sp\sigma$ hopping:
\begin{align} \label{eq:onlyedge}
  l^z_{\bm{\delta}^{\mathrm{i}}\bm{\delta}^{\mathrm{o}}} = \begin{cases}
    \pm i\hbar\sin{\theta} \quad 
    &(\bm{\delta}^{\mathrm{i}}=\bm{R}_0-\bm{R}_{\mp 1},\bm{\delta}^{\mathrm{o}}=\bm{R}_{\pm1}-\bm{R}_{0})\\
    0 \quad 
    &(\mathrm{otherwise})
  \end{cases}.
\end{align}
From the symmetry viewpoint, the imaginary hopping $i(\bm{\delta}^{\rm i}+\bm{\delta}^{\rm o})$ can be interpreted as the activation of the magnetic toroidal dipole $\bm{T}$ corresponding to a time-reversal-odd polar vector, 
while the spin degree of freedom $\bar{\bm{s}}$ can be regarded as the magnetic dipole $\bm{M}$ corresponding to a time-reversal-odd axial vector~\cite{kusunose2022generalization,hayami2024unified}.  
Consequently, $H^{\mathrm{eff}}_{\mathrm{SOI}}$ incorporates the symmetry of the electric dipole $\bm{Q} \, (\sim \bm{T}\times\bm{M})$ corresponding to a time-reversal-even polar vector, 
thereby encoding the polarity associated with the geometric shape. 
Specifically, $\bm{T}$, $\bm{M}$, and $\bm{Q}$ are induced in the $x$, $z$, and $y$ direcitons, respectively, in the present setup.

To calculate the ME tensor using the effective Hamiltonian as Eq.~(\ref{eq:effham}), 
we introduce an effective ME tensor:
\begin{align}\label{eq:anacon}\begin{aligned}
  \alpha^{\nu;\rho}_{\mathrm{eff}}= \left. \alpha^{\nu;\rho}_{\mathrm{eff}}(i\omega_m) \right|_{i\omega_m\rightarrow0+i\gamma}.
\end{aligned}\end{align}
The effective ME tensor defined in terms of the bosonic Matsubara frequencies $\omega_m = 2m\pi k_{\mathrm{B}}T/\hbar$ is expressed by~\cite{Altland_Simons_2010,PhysRevB.110.165111}
\begin{align}\begin{aligned}\label{eq:effsus}
  &\alpha^{\nu;\rho}_{\mathrm{eff}}(i\omega_m)\\
  &\,=- \frac{k_{\mathrm{B}}T}{V} \sum_{i\epsilon_n}
  \Tr \left[ 
    \bar{M}^{\nu}_{\mathrm{eff}} 
    \bar{\mathcal{G}}_{\mathrm{eff}}(i\epsilon_n+i\omega_m)
    \bar{P}^{\rho}
    \bar{\mathcal{G}}_{\mathrm{eff}}(i\epsilon_n)
  \right],
\end{aligned}\end{align}
where $\epsilon_n=(2n+1)\pi k_{\mathrm{B}}T/\hbar$ is the fermionic Matsubara frequencies and $\bar{P}^{\rho}$ is the matrix of polarization operator defined in Eq.~(\ref{eq:pos}).
$\bar{\mathcal{G}}_{\mathrm{eff}}(z)$ is the Green's function of the effective Hamiltonian (the effective Green's function), which is defined as follows:
\begin{align}\label{eq:effgre}
  \bar{\mathcal{G}}_{\mathrm{eff}}(z)=\frac{1}{\hbar z + \mu - \bar{h}_{\mathrm{eff}}(z+\mu/\hbar)},
\end{align}
where $\mu$ is the chemical potential.
The effective Green's function represents the result of incorporating the $p$-orbital contribution into the $s$-orbital Green's function through the $s$--$p$ hybridization~\cite{10.1063/1.4904200}:
\begin{align}\label{eq:geffhyb}
  \bar{\mathcal{G}}_{\mathrm{eff}}(z-\mu/\hbar)
  = \bar{g}_{\parallel}(z) 
  + \bar{g}_{\parallel}(z) \bar{\eta}^{\dagger} \bar{g}_{\perp}(z) \bar{\eta} \bar{g}_{\parallel}(z) + \cdots,
\end{align}
where $\bar{g}_{\parallel}(z)=(\hbar z -  \bar{P}_{s}  \bar{h}  \bar{P}_{s} )^{-1}$.
In order to take into account the contribution of orbital magnetization originating only from the $p$-orbital basis, we
introduce an effective magnetization operator
$\bar{M}^{\nu}_{\mathrm{eff}}=\bar{M}^{\nu}_{\mathrm{eff}}(i\epsilon_n,i\epsilon_n+i\omega_m)$,
\begin{align}\begin{aligned}\label{eq:effMag}
  \bar{M}^{\nu}_{\mathrm{eff}}(z_1,z_2) 
  & =  \bar{P}_{s}  \bar{M}^{\nu}  \bar{P}_{s}  \\
  & \,\,\, + \bar{\eta}^{\dagger} \bar{g}_{\perp}(z_1+\mu/\hbar) \bar{M}^{\nu} \bar{g}_{\perp}(z_2+\mu/\hbar) \bar{\eta}.
\end{aligned}\end{align}

The effective Green's function in Eq.~(\ref{eq:effgre}), the effective magnetization operator in Eq.~(\ref{eq:effMag}), and the effective ME tensor in Eq.~(\ref{eq:effsus}) are illustrated by the Feynman diagrams shown in Fig.~\ref{fig:effsusFD}(a-c), respectively.
For the bubble diagram confined within the $s$-orbital subspace, the effects of the $s$--$p$ hybridization and the magnetic elements defined on the $p$ orbitals are incorporated.
As shown in Eq.~(\ref{eq:geffhyb}) and Fig.~\ref{fig:effsusFD}(a), by including the $s$--$p$ hybridization in the Green's function to infinite order with even numbers of hybridization processes, the effective Green's function $\bar{\mathcal{G}}_{\mathrm{eff}}$ is obtained.
The Green's function components containing an odd number of $s$--$p$ hybridizations propagate through another Green's function, which also includes an odd number of hybridizations mediated by the magnetic operator defined on the $p$-orbitals.
Consequently, the effective ME tensor defined in Eq.~(\ref{eq:effsus}) emerges.

\section{Shape-induced effects in the Green's function} \label{sec:ShapeDip} %-  -  -  -  -  -  -  -  -  -  -  -  -  -  -  -  -  -
We evaluate the angular dependence of the Green's function in Eq.~(\ref{eq:effgre}).
The Hamiltonian matrix can be separated into the $\theta=\pi$ component and the remaining contributions as follows:
\begin{align}\label{eq:SepHam}
  \bar{h}_{\mathrm{eff}} = \left. \bar{h}_{\mathrm{eff}} \right|_{\theta=\pi} + \varDelta\bar{h}_{\mathrm{eff}}^{(\theta)}.
\end{align}
In our effective model in Eq.~(\ref{eq:effham}), $\varDelta\bar{h}_{\mathrm{eff}}^{(\theta)}$ takes a finite value only on the sites adjacent to the apex of the V-shaped 1D chain, as shown in Eq.~(\ref{eq:onlyedge}).
Therefore, $\varDelta\bar{h}_{\mathrm{eff}}^{(\theta)}$ is given by $4\times4$ matrix:
\begin{align}\begin{aligned}\label{eq:dhcomp}
  &\varDelta\bar{h}_{\mathrm{eff}}^{(\theta)}(z) =
  \begin{pmatrix*}
    {0} & 
    {u(z) (1+\cos{\theta}) \bar{\sigma}_0 - i \alpha(z) \sin{\theta} \bar{\sigma}_z} \\
    {(-i \to i)} & 
    {0}
  \end{pmatrix*},
\end{aligned}\end{align}
where the matrix representation corresponds to the two sites adjacent to the apex angle of the V-shaped 1D chain, $\bm{R}_{-1}$ and $\bm{R}_{+1}$, in Eq.~(\ref{eq:kulat}). 
$\bar{\sigma}_0$ and $\bar{\sigma}_{\nu}\,(\nu=x,y,z)$ are the identity matrix and Pauli matrices for spin, respectively.
In this section, $(-i \to i)$ denotes taking the conjugate transpose without taking the conjugate of the coefficients [e.g., $u(z)$ and $\alpha(z)$].
Here, it corresponds to the other off-diagonal component obtained by replacing $-i$ with $i$. 

Then, using $\varDelta\bar{h}_{\mathrm{eff}}^{(\theta)}$, the following Dyson equation is obtained:
\begin{align}\label{eq:Dysontheta}
  \bar{\mathcal{G}}_{\mathrm{eff}} = \bar{\mathcal{G}}_{\mathrm{eff}}^{(\pi)} + \bar{\mathcal{G}}_{\mathrm{eff}}^{(\pi)} \varDelta\bar{h}_{\mathrm{eff}}^{(\theta)} \bar{\mathcal{G}}_{\mathrm{eff}},
\end{align}
where, $\bar{\mathcal{G}}_{\mathrm{eff}}^{(\pi)} = \left. \bar{\mathcal{G}}_{\mathrm{eff}} \right|_{\theta=\pi}.$
It is noted that when $\theta = \pi$, the V-shaped 1D chain represented by Eq.~(\ref{eq:kulat}) corresponds to a simple 1D chain with the spatial inversion symmetry.
Accordingly, $\bar{\mathcal{G}}_{\mathrm{eff}}^{(\pi)}$ is evaluated analytically, with the detailed derivation presented in Appendix~\ref{sec:app:1DChaG}.
Then, by solving Eq.~(\ref{eq:Dysontheta}), we obtain the following expression:
\begin{align}
  \bar{\mathcal{G}}_{\mathrm{eff}}
  = \bar{\mathcal{G}}_{\mathrm{eff}}^{(\pi)}
  + \bar{\mathcal{G}}_{\mathrm{eff}}^{(\pi)} \, \bar{\mathcal{T}}^{(\theta)} \, \bar{\mathcal{G}}_{\mathrm{eff}}^{(\pi)},
\end{align}
where $\bar{\mathcal{T}}^{(\theta)}$ is the transition matrix ($T$-matrix) with respect to the V-shaped geometry:
\begin{align}\label{eq:tmatrix}
  \bar{\mathcal{T}}^{(\theta)}(z) 
  = \varDelta\bar{h}_{\mathrm{eff}}^{(\theta)}
  \frac{1}{1-\bar{\mathcal{G}}_{\mathrm{eff}}^{(\pi)} \, \varDelta\bar{h}_{\mathrm{eff}}^{(\theta)} }.
\end{align}
The deviation from the straight chain is localized around the apex at $n=0$. 
Consequently, the effect of the V-shaped geometry can be represented in a form analogous to the $T$-matrix formalism for a local perturbation, in close analogy to impurity scattering~\cite{PhysRev.79.469,PhysRev.135.A130}.
Therefore, the $T$-matrix $\bar{\mathcal{T}}^{(\theta)}$ encodes information about the shape of the Hamiltonian.
Moreover, since the $T$-matrix reflects the local effects of the shape, it can be analyzed within a smaller subspace than the full Hamiltonian, expressed as the $2(2N+1)\times 2(2N+1)$ matrix.
In our effective model in Eq.~(\ref{eq:effham}), $\bar{\mathcal{T}}^{(\theta)}$ is given by the following $4\times4$ matrix:
\begin{align}\label{eq:Tcomp}
  \bar{\mathcal{T}}^{(\theta)}(z) = \begin{pmatrix}
    {\mathcal{T}_{q_0^{\mathrm{c}}\sigma_0} \bar{\sigma}_0 } & 
    {\mathcal{T}_{q_0^{\mathrm{b}}\sigma_0} \bar{\sigma}_0 - i \mathcal{T}_{t_x^{\mathrm{b}}\sigma_z} \bar{\sigma}_z} \\
    {(-i \to i)} &
    {\mathcal{T}_{q_0^{\mathrm{c}}\sigma_0} \bar{\sigma}_0 } 
  \end{pmatrix}.
\end{align}
$\mathcal{T}_{q_0^{\mathrm{c}}\sigma_0}$, $\mathcal{T}_{q_0^{\mathrm{b}}\sigma_0}$, and $\mathcal{T}_{t_x^{\mathrm{b}}\sigma_z}$ are nonzero components in $\bar{\mathcal{T}}^{(\theta)}$.
To characterize the symmetry properties of the geometry-induced scattering processes, we decompose the $T$-matrix into multipole components, whose coefficients are indicated by the subscripts:
$q_0^{\mathrm{c}}$, $q_0^{\mathrm{b}}$, and $t_x^{\mathrm{b}}$ denote the multipoles defined in the two-site subspace, which will be introduced later in Sec.~\ref{sec:Selection}
and Appendix~\ref{sec:app:MpBME}.
Nonzero off-diagonal components over two sites in $\bar{\mathcal{T}}^{(\theta)}$ are mainly determined by $\varDelta\bar{h}_{\mathrm{eff}}^{(\theta)}$, 
which is composed of the 2nd-nearest-neighbor hopping ($\sim q_0^{\mathrm{b}}\sigma_0$) and the effective spin--orbit interaction (product of imaginary hopping and spin $\sim t_x^{\mathrm{b}}\sigma_z$), as shown in Eq.~(\ref{eq:dhcomp}).
In contrast, the diagonal components are mainly determined by $\bigl(\varDelta\bar{h}_{\mathrm{eff}}^{(\theta)}\bigr)^2$ ($\sim q_0^{\mathrm{c}}\sigma_0$).

Similarly to the Hamiltonian matrix in Eq.~(\ref{eq:SepHam}), the angular $\theta$ dependence of the effective magnetization is separated as follows:
\begin{align}
  \bar{M}^{\nu}_{\mathrm{eff}} = \bar{M}^{\nu(\pi)}_{\mathrm{eff}} + \varDelta \bar{M}^{\nu(\theta)}_{\mathrm{eff}},
\end{align}
where $\bar{M}^{\nu(\pi)}_{\mathrm{eff}} = \left. \bar{M}^{\nu}_{\mathrm{eff}} \right|_{\theta=\pi}$.
In our effective model in Eq.~(\ref{eq:effham}), $\varDelta \bar{M}^{z(\theta)}_{\mathrm{eff}}$ is, like $\bar{\mathcal{T}}^{(\theta)}$, given by the $4\times4$ matrix as follows:
\begin{align}\begin{aligned}\label{eq:dMcomp}
  &\varDelta \bar{M}^{z(\theta)}_{\mathrm{eff}}(z_1,z_2) = \begin{pmatrix}
    {0} & 
    {\varDelta M_{q_0^{\mathrm{b}}\sigma_z} \bar{\sigma}_z - i\varDelta M_{t_x^{\mathrm{b}}\sigma_0} \bar{\sigma}_0} \\
    {(-i \to i)} & 
    {0} 
  \end{pmatrix},
\end{aligned}\end{align}
where $\varDelta M_{q_0^{\mathrm{b}}\sigma_z}$ and $\varDelta M_{t_x^{\mathrm{b}}\sigma_0}$ are nonzero components in $\varDelta \bar{M}^{z(\theta)}_{\mathrm{eff}}$; 
$\varDelta M_{q_0^{\mathrm{b}}\sigma_z}$ corresponds to the contribution from the spin magnetization, while $\varDelta M_{t_x^{\mathrm{b}}\sigma_0}$ corresponds to that from the orbital magnetization.
In our model, $\varDelta \bar{M}^{z(\theta)}_{\mathrm{eff}}$ is constrained to the form given in Eq.~(\ref{eq:dMcomp}) through Eq.~(\ref{eq:effMag}), analogous to $\varDelta\bar{h}_{\mathrm{eff}}^{(\theta)}$ [Eq.~(\ref{eq:dhcomp})].

\begin{figure*}[t]
  \centering
  \includegraphics[width=0.85\linewidth]{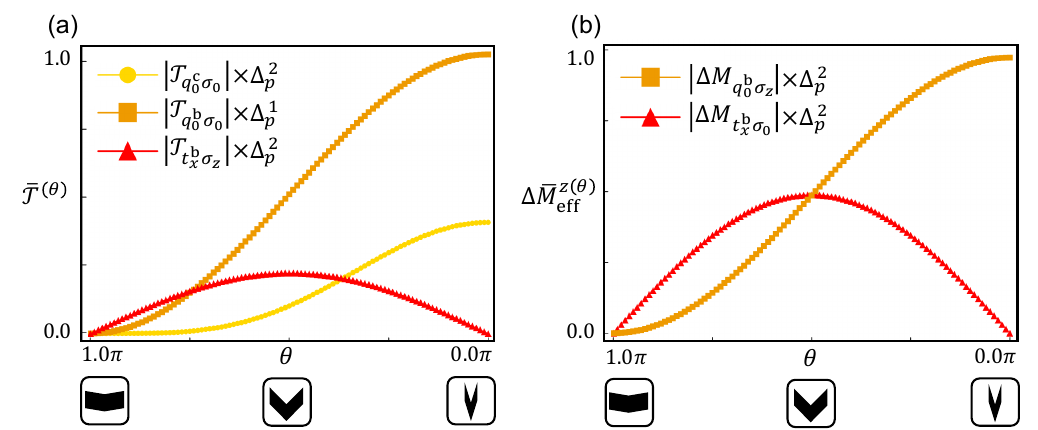}
  \caption{\label{fig:angdipTandM}
  Absolute values of the components of (a) $\bar{\mathcal{T}}^{(\theta)}$ and (b) $\varDelta \bar{M}^{z(\theta)}_{\mathrm{eff}}$ as a function of the apex angle $\theta$.
  We fix the chemical potential $\mu=0.5$ and the complex frequencies $z=z_1=z_2=0.5\pi i$, and we use the other model parameters in Eq.~(\ref{eq:HamPara}) and $N=128$.
  In both cases, the terms including $q_0^{\mathrm{c}}$ and $q_0^{\mathrm{b}}$ exhibit a $(1+\cos{\theta})$-type behavior, whereas that including $t_x^{\mathrm{b}}$ shows a $\sin{\theta}$-type dependence.
  Furthermore, the obtained results display the $\Delta_{p}$ dependence consistent with Eqs.~(\ref{eq:effMag}) and (\ref{eq:Tthdep}).
  }
\end{figure*}

We evaluate the apex angle $\theta$ dependence of $\bar{\mathcal{T}}^{(\theta)}$ and $\varDelta \bar{M}^{z(\theta)}_{\mathrm{eff}}$.
We set the Hamiltonian parameters given in Eq.~(\ref{eq:HamPara}). 
We also set the chemical potential $\mu=0.5$ and the complex frequencies that is an argument of $\bar{\mathcal{T}}^{(\theta)}$ and $\varDelta \bar{M}^{z(\theta)}_{\mathrm{eff}}$: $z=z_1=z_2=0.5\pi i$.
Figure~\ref{fig:angdipTandM}(a) shows the $\theta$ dependences of $\bar{\mathcal{T}}^{(\theta)}$, where we find that each coefficient is expressed by the following expressions as 
\begin{align}\begin{aligned}\label{eq:angdipT}
  \mathcal{T}_{q_0^{\mathrm{c}}\sigma_0} &\propto \Delta_p^{-2} (1+\cos{\theta})^{2}, \\
  \mathcal{T}_{q_0^{\mathrm{b}}\sigma_0} &\propto \Delta_p^{-1} (1+\cos{\theta}), \\
  \mathcal{T}_{t_x^{\mathrm{b}}\sigma_z} &\propto \Delta_p^{-2} \sin{\theta}.
\end{aligned}\end{align}
This dependence is obtained by analyzing the following expression:
\begin{align}\begin{aligned}\label{eq:Tthdep}
  \bar{\mathcal{T}}^{(\theta)}(z) 
  = \varDelta\bar{h}_{\mathrm{eff}}^{(\theta)} 
  + O(\varDelta\bar{h}_{\mathrm{eff}}^{(\theta)}\bar{\mathcal{G}}_{\mathrm{eff}}^{(\pi)}\varDelta\bar{h}_{\mathrm{eff}}^{(\theta)} )
  \quad(\Delta_p\rightarrow\infty).
\end{aligned}\end{align}
Compared to Eq.~(\ref{eq:dhcomp}), 
the off-diagonal components $\mathcal{T}_{q_0^{\mathrm{b}}\sigma_0}$ and $\mathcal{T}_{t_x^{\mathrm{b}}\sigma_z}$ reflect the  $\Delta_p$ and $\theta$ dependence of the 2nd-nearest-neighbor hopping ($u\propto\Delta_p^{-1}$) and the effective spin--orbit interaction ($\alpha\propto\Delta_p^{-2}$), respectively.
Moreover, the diagonal components $\mathcal{T}_{q_0^{\mathrm{c}}\sigma_0}$ arises through the square of the 2nd-nearest-neighbor hopping at the lowest order in $\Delta_p^{-1}$.
It is also noted from the above equation in Eq.~(\ref{eq:Tthdep}) that when  $\bar{\mathcal{G}}_{\mathrm{eff}}^{(\pi)}(z)$ becomes large 
(for example, when $\mathrm{Im}z$ is sufficiently small and $\mathrm{Re}z$ corresponds to the resonant frequency), 
this power expansion breaks down, and the $\theta$ dependence deviates from Eq.~(\ref{eq:angdipT}).

In the same manner, the $\theta$ dependence of $\varDelta \bar{M}^{z(\theta)}_{\mathrm{eff}}$ in Fig.~\ref{fig:angdipTandM}(b)
is given as follows:
\begin{align}\begin{aligned}\label{eq:angdipM}
  \varDelta M_{q_0^{\mathrm{b}}\sigma_z}  &\propto \Delta_p^{-2} (1+\cos{\theta}), \\
  \varDelta M_{t_x^{\mathrm{b}}\sigma_0}  &\propto \Delta_p^{-2} \sin{\theta}.
\end{aligned}\end{align}
This behavior is understood from Eq.~(\ref{eq:effMag});
$\bar{g}_{\perp}(z)\propto\Delta_p^{-1}(\Delta_p\gg z)$ at the lowest order in $\Delta_p^{-1}$.
$\varDelta M_{q_0^{\mathrm{b}}\sigma_z}$ and $\varDelta M_{t_x^{\mathrm{b}}\sigma_0}$ arise from the even- and odd-order terms of the angular momentum operator $\bar{l}^{z}$, as given in Eqs.~(\ref{eq:2NNHdep}) and (\ref{eq:eSOIdep}), respectively.
Therefore, the resulting $\theta$ dependence is analogous to that of $\varDelta\bar{h}_{\mathrm{eff}}^{(\theta)}$ shown in Eq.~(\ref{eq:dhcomp}). 
Furthermore, because the angular dependence originates solely from $\bar{\eta}$, no $\theta$ dependence appears in the denominator term; 
consequently, Eq.~(\ref{eq:angdipM}) appears as a direct proportional relation, unlike $\bar{\mathcal{T}}^{(\theta)}$.

\section{Multipole-Based Selection Rules} \label{sec:Selection} %-  -  -  -  -  -  -  -  -  -  -  -  -  -  -  -  -  -

Finally, we introduce the multipole basis and clarify the underlying mechanism by taking into account the selection rules.
By employing the symmetry-adapted multipole basis~\cite{kusunose2022generalization, Hayami_PhysRevLett.122.147602, PhysRevB.102.144441, doi:10.7566/JPSJ.89.104704, doi:10.7566/JPSJ.91.014701, Kusunose_PhysRevB.107.195118, hayami2024unified, 2s5q-p42x}, 
arbitrary operators such as the effective Hamiltonian can be expanded by the products of the cluster and bond degrees of freedom (denoted as cluster multipole and bond multipole, respectively) and the spin degree of freedom.

Table~\ref{tab:MpBofOperator} summarizes the multipole degrees of freedom included in each operator in our effective model.
The onsite degrees of freedom are described by the cluster electric multipole $\bar{q}^{\mathrm{c}}$, 
while the off-site degrees of freedom are described by the bond electric multipole $\bar{q}^{\mathrm{b}n}$ and the bond magnetic toroidal multipole $\bar{t}^{\mathrm{b}n}$; the former (latter) corresponds to the real (imaginary) hopping. 
Here, the superscript $n$ denotes the bond between the $n$th nearest-neighbor sites, and the prime (${}^{\prime}$) indicates the quantity defined only on the two sites adjacent to the apex angle ($\bm{R}_{-1}$ and $\bm{R}_{+1}$) of the V-shaped 1D chain.
The spin degrees of freedom are represented by the Pauli matrices: $\bar{\sigma}_0$ corresponds to the electric monopole, while $(\bar{\sigma}_x,\bar{\sigma}_y,\bar{\sigma}_z)$ represents the magnetic dipole.
The matrix representations of the relevant multipoles are given in Appendix~\ref{sec:app:MpB}.

\begin{table*}[t]
  \centering
  \caption{\label{tab:MpBofOperator}
    List of multipoles included in the representative operators related to the ME tensor.
    For simplicity, notations such as ``\,$\bar{~}$\,'' are omitted
    (e.g., ${\mathcal{G}}^{(\pi)}=\bar{{\mathcal{G}}}^{(\pi)}_{\mathrm{eff}}$).
    In the multipole notation, the superscripts $\mathrm{c}$ and $\mathrm{b}$ denote cluster and bond multipoles, the superscript $n$ following $\mathrm{b}$ denotes the $n$th nearest neighbor, 
    the subscripts $0$ and $(x,y)$ denote the monopole and dipole, respectively, and the prime (${}^{\prime}$) indicates the quantity defined on the two sites adjacent to the apex angle ($\bm{R}_{-1}$ and $\bm{R}_{+1}$) of the V-shaped 1D chain. 
    For $\bar{M}^{z(\pi)}_{\mathrm{eff}}$, the edge correction due to the 2nd-nearest-neighbor hopping to the onsite term is not taken into account.
    In the second column, the corresponding irreducible representations under the point group $C_{\mathrm{2v}}$ are shown, where the superscript $\pm$ denotes the parity with respect to the time-reversal operation.
    In the last column, the dependence on the apex angle $\theta$ and the $s$--$p$ energy separation $\Delta_p\,(\Delta_p\rightarrow\infty)$ is shown. 
  }
  \begingroup
  \renewcommand{\arraystretch}{1.55}
  \begin{tabular*}{0.75\linewidth}{@{\extracolsep{\fill}} cccc } % arXiv
    \hline \hline
    Operator & Irreducible representation & Relevant multipole & $\theta,\Delta_p$ dep. \\
    \hline
    $ {\mathcal{G}}^{(\pi)} $    & $ \mathrm{A}_{1}^+ $ & 
    $ q_0^{\mathrm{c}}\sigma_0 ,\, q_0^{\mathrm{b}1}\sigma_0,\,q_0^{\mathrm{b}2}\sigma_0,\ldots $ & $ \propto 1$  \\

    $ {M}^{z(\pi)} $             & $ \mathrm{B}_{1}^- $ & 
    $ q_0^{\mathrm{c}}\sigma_z $ & $ \propto 1 $  \\
    $ $ & $ $ &
    $ q_0^{\mathrm{b}2}\sigma_z $ & $ \propto \Delta_p^{-2} $ \\

    $ {\mathcal{T}}^{(\theta)} $ & $ \mathrm{A}_{1}^+ $ & 
    $ q_0^{\mathrm{c}\prime}\sigma_0 $ & $ \sim \Delta_p^{-2} (1+\cos{\theta})^{2}$ \\
    $ $ & $ $ &
    $ q_0^{\mathrm{b}2\prime}\sigma_0 $ & $ \sim \Delta_p^{-1} (1+\cos{\theta})$ \\
    $ $ & $ $ &
    $ t_x^{\mathrm{b}2\prime}\sigma_z $ & $ \sim \Delta_p^{-2} \sin{\theta}$ \\
    
    $ \varDelta {M}^{z(\theta)} $   & $ \mathrm{B}_{1}^- $ & 
    $ q_0^{\mathrm{b}2\prime}\sigma_z $ & $ \propto \Delta_p^{-2} (1+\cos{\theta})$\\
    $ $ & $ $ &
    $ t_x^{\mathrm{b }2\prime}\sigma_0 $ & $ \propto \Delta_p^{-2} \sin{\theta}$ \\
    
    $ {P}^{x} $                  & $ \mathrm{B}_{1}^+ $ & 
    $ q_x^{\mathrm{c}}\sigma_0 $ & $ \propto \sin{\theta/2}$\\
    
    $ {P}^{y} $                  & $ \mathrm{A}_{1}^+ $ & 
    $ q_y^{\mathrm{c}}\sigma_0 $ & $ \propto \cos{\theta/2}$\\
    \hline \hline
  \end{tabular*}
  \endgroup
\end{table*}

Table~\ref{tab:MpBofOperator} presents the dependence on the apex angle $\theta$ and the $s$--$p$ energy separation $\Delta_p$ for $\Delta_p\rightarrow\infty$.
The simple 1D chain Green's function $\bar{\mathcal{G}}_{\mathrm{eff}}^{(\pi)}$ includes all cluster and bond multipoles generated by arbitrary powers of the hopping Hamiltonian.
The simple 1D chain effective magnetization $\bar{M}_{\mathrm{eff}}^{z(\pi)}$ contains $\bar{q}_0^{\mathrm{c}}\bar{\sigma}_z$ component and also has $\bar{q}_0^{\mathrm{b}2}\bar{\sigma}_z$ component associated with the 2nd-nearest-neighbor hopping. 
Both originate from the spin magnetization contribution.
$\bar{\mathcal{T}}^{(\theta)}$ and $\varDelta \bar{M}^{z(\theta)}_{\mathrm{eff}}$ are discussed in Sec.~\ref{sec:ShapeDip}.
It is noteworthy that, once the effective magnetization is introduced, the magnetic dipole operator is no longer a pure magnetic dipole but acquires an admixture of the magnetic toroidal dipole $\bar{t}_x^{\mathrm{b}2\prime}\bar{\sigma}_0$.
Although the magnetic dipole and the magnetic toroidal dipole are distinct multipoles, they belong to the same irreducible representation in the present system. 
The multipole $\bar{t}_x^{\mathrm{b}2\prime}\bar{\sigma}_0$ arises from the coupling between the orbital magnetization and the $sp\sigma$ hopping. 
Finally, the $\theta$ dependence of $\bar{P}^{x}$ and $\bar{P}^{y}$ is given from Eq.~(\ref{eq:kulat}).

In order for the ME tensor defined in Eq.~(\ref{eq:effsus}) to acquire a finite value, the condition $\Tr[\ast] \ne 0$ must be satisfied.
Within the framework of the multipole basis, the following relations play a key role:
\begin{subequations}\label{eq:selection}\begin{align}
  \bar{q}_0^{\mathrm{b}2\prime} \, \bar{t}_x^{\mathrm{b}2\prime} \, \bar{q}_x^{\mathrm{c}} &= -i \, \bar{q}_0^{\mathrm{c}\prime},\\
  \bar{\sigma}_z \, \bar{\sigma}_z &= \bar{\sigma}_0.
\end{align}\end{subequations}
From the symmetry viewpoint, these equation represent $\mathrm{A}_1^{+}\otimes\mathrm{B}_1^{-}\otimes\mathrm{B}_1^{+}=\mathrm{A}_1^{-}$ and $\mathrm{B}_1^{-}\otimes\mathrm{B}_1^{-}=\mathrm{A}_1^{+}$, respectively.
By taking into account the above processes and extracting the contributions that satisfy Eq.~(\ref{eq:selection}) while retaining only the terms in powers of $\Delta_p^{-1}$, 
the dominant contributions can be expressed as
\begin{align}
  \alpha^{z;x}_{\mathrm{eff}}(i\omega_m)
  =\alpha^{z;x}_{\mathrm{i}}+\alpha^{z;x}_{\mathrm{ii}}+\alpha^{z;x}_{\mathrm{iii}}+\alpha^{z;x}_{\mathrm{iv}}
  +O(\Delta_p^{-4})
\end{align}
The corresponding diagrams are schematically shown in Fig.~\ref{fig:MEprocess}.
The detailed calculations and the explicit expressions are provided in Appendix~\ref{sec:app:MpBME}.
From Table~\ref{tab:MpBofOperator}, the $\theta$ and $\Delta_p$ dependence of $\alpha^{z;x}_{\mathrm{i}}$, $\alpha^{z;x}_{\mathrm{ii}}$, $\alpha^{z;x}_{\mathrm{iii}}$, and $\alpha^{z;x}_{\mathrm{iv}}$ is given as follows:
\begin{subequations}\label{eq:MEcontriddep}\begin{align}
  \alpha^{z;x}_{\mathrm{i}}   & \propto \Delta_p^{-2} \sin\theta \sin{\frac{\theta}{2}},\\
  \alpha^{z;x}_{\mathrm{ii}}  & \sim    \Delta_p^{-3} (1+\cos{\theta}) \sin\theta \sin{\frac{\theta}{2}},\\
  \alpha^{z;x}_{\mathrm{iii}} & \sim    \Delta_p^{-2} \sin\theta \sin{\frac{\theta}{2}},\\
  \alpha^{z;x}_{\mathrm{iv}}  & \sim    \Delta_p^{-3} (1+\cos{\theta}) \sin\theta \sin{\frac{\theta}{2}}.
\end{align}\end{subequations}
As illustrated in Fig.~\ref{fig:MEprocess}, the ME response in our model arises from two distinct processes.
The components $\alpha^{z;x}_{\mathrm{i}}$ and $\alpha^{z;x}_{\mathrm{ii}}$ are represented by the coupling among the polarization ($P_{q_x^{\mathrm{c}} \sigma_0}$) and the magnetization ($\varDelta M_{t_x^{\mathrm{b}}\sigma_0}$) through the hopping across the apex angle ($\bar{q}_0^{\mathrm{b}2}\bar{\sigma}_0$) in the Green's function, 
where $P_{q_x^{\mathrm{c}} \sigma_0}$ and $\varDelta M_{t_x^{\mathrm{b}}\sigma_0}$ represent the charge-potential gradient induced by the applied electric field and the orbital magnetization, respectively.
Thus, $\alpha^{z;x}_{\mathrm{i}}$ ($\alpha^{z;x}_{\mathrm{ii}}$) represents the (higher-order) contributions arising from the coupling between charge-potential gradient and the orbital magnetization.
In contrast, the components $\alpha^{z;x}_{\mathrm{iii}}$ and $\alpha^{z;x}_{\mathrm{iv}}$ are represented by the coupling among the polarization ($P_{q_x^{\mathrm{c}} \sigma_0}$), the spin magnetization ($M_{q_0^{\mathrm{c}}\sigma_z}$), and the spin-dependent hopping across the apex angle in the $T$-matrix ($\bar{t}_x^{\mathrm{b}2\prime}\bar{\sigma}_z$).
Since $\bar{t}_x^{\mathrm{b}2\prime}\bar{\sigma}_z$ represents the contribution from the effective spin--orbit interaction [Eq.~(\ref{eq:Tthdep})], $\alpha^{z;x}_{\mathrm{iii}}$ ($\alpha^{z;x}_{\mathrm{iv}}$) is interpreted as the (higher-order) couplings between the charge-potential gradient and spin magnetizations via the effective spin--orbit interaction.

\begin{figure*}[t]
  \centering
  \includegraphics[width=\linewidth]{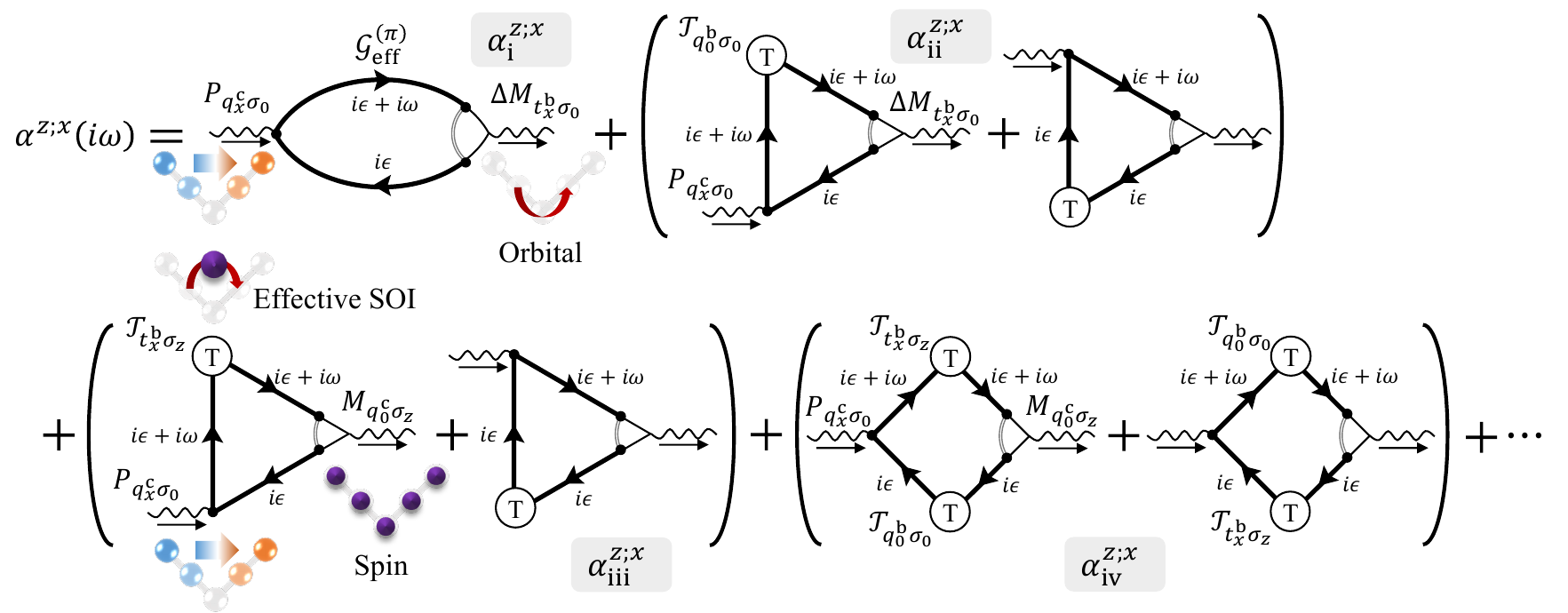}
  \caption{\label{fig:MEprocess}
  Diagrams representing distinct processes for the ME effect in our effective models.
  The thick black line represents the simple 1D chain Green's function at $\theta=\pi$, 
  and the circled T ({\raise0.2ex\hbox{\textcircled{\scriptsize{T}}}}) represents the $T$-matrix about the V shape in Eq.~(\ref{eq:tmatrix}).
  The subscripts of the Matsubara frequencies are omitted.
  The symbols (e.g., $P_{q_x^{\mathrm{c}}\sigma_0}$) attached to the operators correspond to their multipole components in Table~\ref{tab:MpBofOperator}.
  The physical mechanisms corresponding to each multipole component are shown along with their schematic illustrations (see also Fig.~\ref{fig:MpB} in Appendix~\ref{sec:app:MpB}).
  There are two primary physical mechanisms contributing to the ME effect:
  One is the coupling between the charge-potential gradient induced by the incident electric field and the orbital magnetization ($\alpha^{z;x}_{\mathrm{i}}$ and $\alpha^{z;x}_{\mathrm{ii}}$).
  The other is the spin magnetizations induced by the electric field through the effective spin--orbit interaction ($\alpha^{z;x}_{\mathrm{iii}}$ and $\alpha^{z;x}_{\mathrm{iv}}$).
  }
\end{figure*}

\begin{figure}[b]
  \centering
  \includegraphics[width=1.0\linewidth]{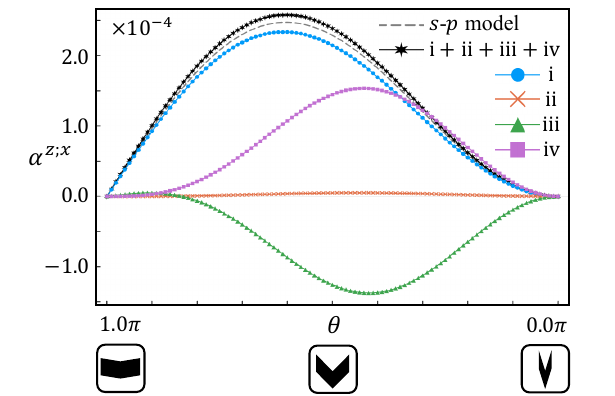} % arXiv
  \caption{\label{fig:MEcontri}
    Apex angle $\theta$ dependence of the components of the effective ME tensor $\alpha^{z;x}_{\mathrm{i}},\alpha^{z;x}_{\mathrm{ii}},\alpha^{z;x}_{\mathrm{iii}}$, and $\alpha^{z;x}_{\mathrm{iv}}$.
    We also present their total contribution and the result calculated using the Kubo formula [Eq.~(\ref{eq:KuboFml})] in the $s$--$p$ model.
    We use the same model parameters as those in Fig.~\ref{fig:Qysp}(a).
    For the analytic continuation, we employed the barycentric rational function approximation method~\cite{PhysRevB.111.125139}.
  }
\end{figure}

Figure~\ref{fig:MEcontri} shows the $\theta$ dependence of the components $\alpha^{z;x}_{\mathrm{i}},\alpha^{z;x}_{\mathrm{ii}},\alpha^{z;x}_{\mathrm{iii}}$, and $\alpha^{z;x}_{\mathrm{iv}}$ at the chemical potential $\mu=0.5$, where we use the same model parameters as those in Fig.~\ref{fig:Qysp}(a).
For the analytic continuation from $i\omega_m$ to $0+i\gamma$, we employed the barycentric rational function approximation method~\cite{PhysRevB.111.125139}.
Here, the fermionic Matsubara frequencies were summed from $n=-100$ to $100$, while the bosonic Matsubara frequency data were taken over the range $m=0$ to $30$.
We also present their total contribution and the result calculated by using the Kubo formula [Eq.~(\ref{eq:KuboFml})] in the original $s$--$p$ model.
It is demonstrated that the summation of the above four components successfully reproduces the result obtained from the $s$--$p$ model, indicating the dominant contributions are given by $\alpha^{z;x}_{\mathrm{i}}$--$\alpha^{z;x}_{\mathrm{iv}}$.
Among them, we find that the contribution from $\alpha^{z;x}_{\mathrm{i}}$ is the most dominant, and thereby, the $\theta$ dependence of the total $\alpha^{z;x}_{\rm eff}$ is governed as
\begin{align}
  \alpha^{z;x}_{\mathrm{eff}} \propto \sin\theta \sin{\frac{\theta}{2}}.
\end{align}
This behavior arises from the competition between the maximization of the orbital angular momentum at the angle of the V shape ($\propto \sin \theta$) and the gain in the charge-potential gradient induced by the electric field along the $x$ direction ($\propto \sin \theta/2$).
It exhibits a peak at $\theta = 2 \tan^{-1}\sqrt{2} \approx 0.608\pi$, which is consitent with the result in Fig.~\ref{fig:Qysp}(b).

\section{Ferromagnetic System} \label{sec:Ferromag} %-  -  -  -  -  -  -  -  -  -  -  -  -  -  -  -  -  -

\begin{figure*}[t]
  \centering
  \includegraphics[width=0.85\linewidth]{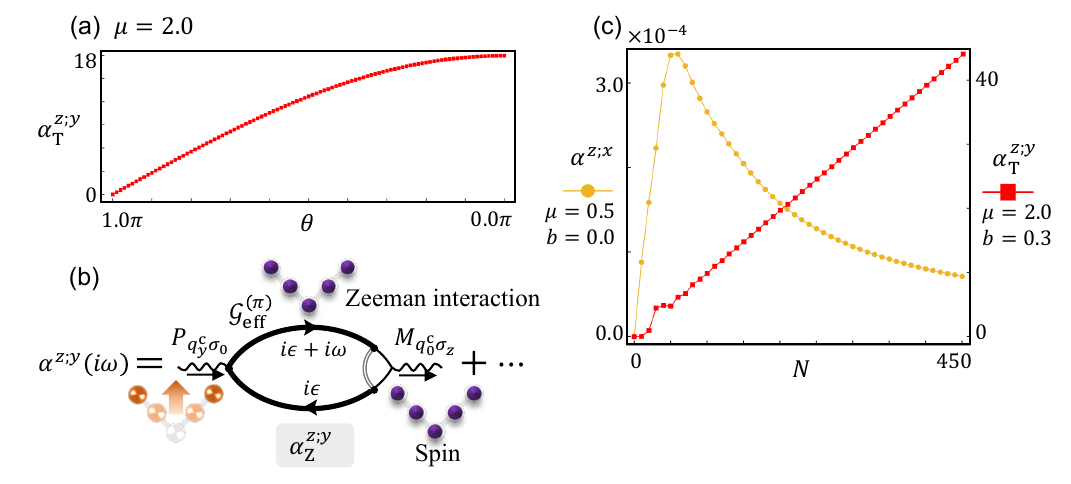}
  \caption{\label{fig:ferromag}
  (a) 
  Apex angle $\theta$ dependence of the isothermal ME tensor $\alpha_{\mathrm{T}}^{z;y}$ at $\mu=2$ in the ferromagnetic $s$--$p$ model.
  We set the Hamiltonian parameters as Eq.~(\ref{eq:HamPara}), and $N=128$ and $T=0.01$.
  The ME response increases monotonically as the apex angle $\theta$ decreases.
  (b) 
  Diagrams for the ME effect in the ferromagnetic V-shaped model.
  (c)
  The V-shaped lattice size $N$ dependence of the nonmangetic-driven ME tensor $\alpha^{z;x}$ ($\mu=0.5,b=0$) and the magnetic-driven ME tensor $\alpha^{z;y}_{\mathrm{T}}$ ($\mu=2,b=0.3$) at $\theta=\pi/2$.
  The difference in physical mechanisms results in different size dependence.
  }
\end{figure*}

In this section, we consider the effect of the time-reversal-symmetry breaking by additionally taking into account the ferromagnetic mean field along the $z$ direction. 
For that purpose, we include a Zeeman interaction in the Hamiltonian:
\begin{align}
  H_{\mathrm{Zee}} 
  &= \frac{b}{\hbar} \left( L^{z} + \mathrm{g} S^{z} \right).
\end{align}
The introduction of a ferromagnetic moment along the $z$ direction induces a magnetic toroidal dipole $T_x$ along the $x$ direction through its coupling to the electric polarization along the $y$ direction, which arises from the V-shaped geometry.
Consequently, from the viewpoint of symmetry, the ME response driven by the magnetic toroidal dipole, i.e., the magnetic-driven ME, is induced in the $\alpha^{z;y}$ component~\cite{PhysRevB.98.165110,PhysRevB.104.054412,hayami2024unified}.

We investigate the shape dependence of the isothermal ME tensor $\alpha_{\mathrm{T}}^{z;y}=\left. \alpha^{z;y} \right|_{\gamma=0}$ in the V-shaped $s$--$p$ model.
Figure~\ref{fig:ferromag}(a) shows the $\theta$ dependence of the isothermal ME response.
We use the Hamiltonian parameters in Eq.~(\ref{eq:HamPara}) and $b=0.3$; the other parameters are set to $N=128,\mu=2$, and $T=0.01$.
In contrast to the previous results of nonmagnetic-driven $\alpha^{z;x}$, the magnitude increases monotonically as the apex angle $\theta$ decreases. 

In the magnetic-driven ME response, the underlying mechanism is more straightforward than that of the nonmagnetic-driven $\alpha^{z;x}$, since the Zeeman interaction introduces an additional component into $\bar{\mathcal{G}}^{(\pi)}$ as $\bar{q}_0^{\mathrm{c}}\bar{\sigma}_z$, which leads to a natural expansion as follows:
\begin{align}\label{eq:METxSepa}
  \alpha_{\mathrm{eff}}^{z;y}(i\omega_m)&=\alpha_{\mathrm{Z}}^{z;y}+O(\Delta_p^{-1}).
\end{align}
$\alpha_{\mathrm{T}}^{z;y}$ and $\alpha_{\mathrm{eff}}^{z;y}(i\omega_m)$ are related by Eq.~(\ref{eq:anacon}).
The corresponding diagram is shown in Fig.~\ref{fig:ferromag}(b), and its explicit expressions are provided in Appendix~\ref{sec:app:MpBME}.
The diagram in Fig.~\ref{fig:ferromag}(b) displays that $\alpha_{\mathrm{T}}^{z;y}$ originates from the spin magnetization induced by the coupling between the charge-potential gradient along the $y$ direction generated by the applied electric field and the Zeeman interaction. 
Its $\theta$ dependence is given by
\begin{align}\label{eq:MEzeeddep}
  \alpha_{\mathrm{Z}}^{z;y}&\propto \cos{\frac{\theta}{2}}.
\end{align}
Accordingly, the angular dependence obtained in Fig.~\ref{fig:ferromag}(a) can be understood as a result of the increased gain of the charge-potential gradient along the $y$ direction when the apex angle becomes sharper. 

In addition, the difference in physical mechanisms between the magnetic- and nonmagnetic-driven ME responses results in distinct size dependence.
Figure~\ref{fig:ferromag}(c) shows the V-shaped lattice size $N$ dependence of the nonmagnetic-driven ME tensor $\alpha^{z;x}$ and the ferromagnetic-driven ME tensor $\alpha^{z;y}_{\mathrm{T}}$ in the $s$--$p$ model.
The former $\alpha^{z;x}$, originating from local orbital magnetization and effective spin--orbit interaction, decreases with increasing system size.
In contrast, the latter $\alpha^{z;y}_{\mathrm{T}}$, which arises from the coupling between the charge-potential gradient and the Zeeman interaction acting over the entire system, increases proportionally with system size.
Taking into account that a constant electric field is applied, this calculation of the magnetic-driven ME effect demonstrates that the induced magnetization is proportional to the applied voltage along the $y$ direction.

\section{Summary} \label{sec:Summary} %-------------------------------------------------------------------------------------

In summary, we have theoretically investigated the shape dependence of both nonmagnetic- and magnetic-driven ME responses in the noncentrosymmetric V-shaped chain model.
The key advances of this work are threefold: 
(i) the identification of the characteristic angular dependence of the ME response originating from the interplay between orbital motion and electric polarization in the V-shaped geometry,
(ii) the characterization of local geometric effects through a scattering-theoretical analysis, and 
(iii) the identification of the dominant microscopic contributions to the ME response using the multipole representation.

To demonstrate these points,  we first evaluated the shape dependence of the nonmagnetic-driven ME effect numerically based on the Kubo formula in a model including $s$ and $p$ orbitals. 
As a result, the magnitude of the ME effect reaches its maximum at an apex angle of $\theta \approx 0.6 \pi$.
To clarify the origin of the shape dependence, we subsequently performed a detailed theoretical analysis of the numerical results.
We then constructed a low-energy effective Hamiltonian by projecting onto the $s$-orbital basis. 
Consequently, we found that the polarity induced by the V-shaped geometry manifests as an effective spin--orbit interaction.
Furthermore, by analyzing the shape dependence of the Green's function, we clarified that the geometric effect is described by a $T$-matrix contribution to the Green's function.

Next, we analyzed the selection rules using the multipole basis.
It was revealed that the effective spin--orbit interaction and the orbital angular momentum generated by the motion across the apex angle play an essential role in the ME effect.
We also found that the term with angle dependence of  $\sin{\theta} \sin{\theta/2}$, which exhibits a maximum at $\theta = 2 \tan^{-1}\sqrt{2} \approx 0.608\pi$, is dominant.

Finally, we also investigated the V-shaped ferromagnetic model including the Zeeman interaction.
We found that the magnetic-driven ME response originates from the spin magnetization induced by the coupling between the charge-potential gradient along the $y$ direction generated by the applied electric field and the Zeeman interaction. 
The difference in physical mechanisms results in distinct shape and size dependence.

Our results clarify the microscopic mechanisms underlying the shape-dependent ME responses in the finite V-shaped chain. 
In particular, the scattering-theoretical analysis allows us to characterize the effects of local geometric symmetry breaking in terms of the $T$-matrix, while the multipole representation helps identify the dominant microscopic contributions to the response.
The predicted angular dependence may be experimentally examined in suitable V-shaped nanostructures.

These findings highlight the potential of nanoscale geometric design for controlling electronic responses through local symmetry breaking and effective spin–orbit interactions. 
In realistic nano- and mesoscopic structures, additional effects, such as plasmonic excitations and local electromagnetic fields, may also play important roles in determining their response properties~\cite{Link1999,Jain2006,Billot15,Babynina2016}. 
Investigating the interplay between these effects and the microscopic mechanisms identified in this work is an interesting direction for future research.

Furthermore, the microscopic mechanism of the ME effect discussed in this work may provide insights into bulk systems such as zigzag chain systems, 
where local inversion symmetry is broken and a variety of odd-parity multipole physics are expected~\cite{doi:10.7566/JPSJ.83.014703,Hayami_doi:10.7566/JPSJ.84.064717, Hayami_doi:10.7566/JPSJ.85.053705, PhysRevB.105.155157, Suzuki_PhysRevB.105.075201, m96k-8ndj}
(e.g., 
CeRu$_2$Al$_{10}$~\cite{Tursina:wm6046,doi:10.1143/JPSJ.80.073701}, 
Ce$_3$TiBi$_5$~\cite{doi:10.7566/JPSCP.30.011189,doi:10.7566/JPSJ.89.033703}, 
$\alpha$-YbAl$_{1-x}$Mn$_x$B$_4$~\cite{PhysRevResearch.3.023140}, and 
NdRu$_2$Al$_{10}$~\cite{sudo2025large}
).
Although the present analysis focuses on a finite V-shaped chain, further investigation is required to clarify its applicability to periodic bulk systems.

%- acknowledgments ---------------------------------------------------------------------------------------------------------
\begin{acknowledgments}
This research was supported by 
JST SPRING (Grant No.~JPMJSP2119),
JSL BOOST (Grant No.~JPMJBS2426),
JSPS KAKENHI (Grants No. JP22H00101, No. JP22H01183, No. JP23H04869, and No. JP23K03288), 
JST CREST (Grant No.~JPMJCR23O4), 
and JST FOREST (Grant No.~JPMJFR2366).

\end{acknowledgments}

%- Data Availability -----------------------------------------------------------------------------------------------------------------------
\section*{Data Availability}
The data that support the findings of this article are openly available in Zenodo at \cite{dataset_zenodo}.

%- appendix ----------------------------------------------------------------------------------------------------------------
\appendix

\newpage % arXiv
\begin{widetext}
\newpage % arXiv
\section{Green's function of simple 1D chain } \label{sec:app:1DChaG} %--------------------------------------------------------------------------------------------
When $\theta = \pi$, the V-shaped 1D chain represented by Eq.~(\ref{eq:kulat}) corresponds to a simple 1D chain.
Here, we consider the nonmagnetic case for simplicity. 
The Hamiltonian matrix 
\begin{align}
  \bar{h}^{(\pi)}=\left. \bar{h}_{\mathrm{eff}}(z+\mu/\hbar) \right|_{\theta=\pi}
\end{align}
can be expressed as $(2N+1)\times(2N+1)$ matrix for the $2N+1$ sites except for the spin degree of freedom:
\begin{align}
  \bar{h}^{(\pi)} &= \bar{h}_0 + \bar{h}_{\mathrm{edge}},
\end{align}
\begin{align}
  \bar{h}_0 = \begin{pmatrix}
    {\varepsilon_0-t_2} & {t_1} & {t_2} & {0} & {0} & \dots  & {0} & {0} & {0} \\
    {t_1} & {\varepsilon_0} & {t_1} & {t_2} & {0} & \dots  & {0} & {0} & {0} \\
    {t_2} & {t_1} & {\varepsilon_0} & {t_1} & {t_2} & \dots  & {0} & {0} & {0} \\
    {0} & {t_2} & {t_1} & {\varepsilon_0} & {t_1} & \dots  & {0} & {0} & {0} \\
    {0} & {0} & {t_2} & {t_1} & {\varepsilon_0} & \dots  & {0} & {0} & {0} \\
    \vdots & \vdots & \vdots & \vdots & \vdots & \ddots & \vdots & \vdots & \vdots\\
    {0} & {0} & {0} & {0} & {0} & \dots  & {\varepsilon_0} & {t_1} & {t_2} \\
    {0} & {0} & {0} & {0} & {0} & \dots  & {t_1} & {\varepsilon_0} & {t_1} \\
    {0} & {0} & {0} & {0} & {0} & \dots  & {t_2} & {t_1} & {\varepsilon_0-t_2} \\
  \end{pmatrix},\quad
  \bar{h}_{\mathrm{edge}} = \begin{pmatrix}
    {2t_2} & {0} & {0}  & \dots  & {0} & {0} & {0} \\
    {0} & {0} & {0} &  \dots  & {0} & {0} & {0} \\
    {0} & {0} & {0} &  \dots  & {0} & {0} & {0} \\
    \vdots & \vdots & \vdots  & \ddots & \vdots & \vdots & \vdots \\
    {0} & {0} & {0} &  \dots  & {0} & {0} & {0} \\
    {0} & {0} & {0} &  \dots  & {0} & {0} & {0} \\
    {0} & {0} & {0} &  \dots  & {0} & {0} & {2t_2} \\
  \end{pmatrix}, 
\end{align}
where
\begin{align}
  \varepsilon_0=2u(z+\mu/\hbar), \quad
  t_1=-t_{ss\sigma}, \quad
  t_2=-u(z+\mu/\hbar).
\end{align}
By diagonalizing $\bar{h}_0$, the eigenenergies $E^{(0)}_{k_m}$ and the eigenstates $ \psi^{(0)}_{k_m}(R_n^x) $ are obtained as follows:
\begin{subequations}\begin{align}
  E^{(0)}_{k_m} 
  &= \varepsilon_0 + 2t_1\cos(k_m a ) + 2 t_2\cos(2k_m a ),\\
  \psi^{(0)}_{k_m}(R_n^x)
  &= \begin{cases}
    \displaystyle \sqrt{\frac{2 a }{L}}\sin(k_m R_n^x) \quad &(m=\mathrm{even})\\
    \displaystyle \sqrt{\frac{2 a }{L}}\cos(k_m R_n^x) \quad &(m=\mathrm{odd} )\\
  \end{cases}
\end{align}\end{subequations}
where $L=(2N+2)a$,
\begin{align}
  k_m = \frac{\pi}{L}m \quad (m=1,2,...,2N+1).
\end{align}
Therefore, the Green's function of $\bar{h}_0$ is given by as follows:
\begin{align}
  \left[\bar{\mathcal{G}}_0(z)\right]^{nn'}_{\sigma\sigma'}=\sum_{k_m}\frac{\psi^{(0)}_{k_m}(R^x_{n}) \, \psi^{(0)\ast}_{k_m}(R^x_{n'})}{\hbar z + \mu - E^{(0)}_{k_m} } \left[\bar{\sigma}_0\right]_{\sigma\sigma'}.
\end{align}
Then, the Green's function of $\bar{h}^{(\pi)}$ is obtained from the Dyson equation:
\begin{align}
  \bar{\mathcal{G}}_{\mathrm{eff}}^{(\pi)}
  = \bar{\mathcal{G}}_0 
  + \bar{\mathcal{G}}_0 \bar{h}_{\mathrm{edge}}\bar{\mathcal{G}}_{\mathrm{eff}}^{(\pi)}.
\end{align}
Solving this equation, we obtain
\begin{align}
  \bar{\mathcal{G}}_{\mathrm{eff}}^{(\pi)} 
  = \bar{\mathcal{G}}_0
  + \bar{\mathcal{G}}_0\bar{h}_{\mathrm{edge}} \frac{1}{1-\bar{\mathcal{G}}_0\bar{h}_{\mathrm{edge}}} \bar{\mathcal{G}}_0.
\end{align}

\newpage % arXiv
\section{Matrix representation of multipoles} \label{sec:app:MpB} %--------------------------------------------------------------------------------------------

We show the matrix elements of the multipole operators~\cite{Hayami_PhysRevLett.122.147602, PhysRevB.102.144441,doi:10.7566/JPSJ.91.014701,hayami2024unified}.
The onsite and real bond degrees of freedom are described by electric multipoles $\bar{q}^{\mathrm{c}}$ and $\bar{q}^{\mathrm{b}n}$ and the imaginary bond degrees of freedom are described by magnetic toroidal multipoles $\bar{t}^{\mathrm{b}n}$.
As an example, for the $7$-site basis ($N=3$), the relevant matrix elements of the multipoles are given as follows:
\begin{align}\begin{aligned}
  \bar{q}_0^{\mathrm{c}} &= \begin{pmatrix}
    {1} & {0} & {0} & {0} & {0} & {0} & {0} \\
    {0} & {1} & {0} & {0} & {0} & {0} & {0} \\
    {0} & {0} & {1} & {0} & {0} & {0} & {0} \\
    {0} & {0} & {0} & {1} & {0} & {0} & {0} \\
    {0} & {0} & {0} & {0} & {1} & {0} & {0} \\
    {0} & {0} & {0} & {0} & {0} & {1} & {0} \\
    {0} & {0} & {0} & {0} & {0} & {0} & {1} \\
  \end{pmatrix},\,
  \bar{q}_0^{\mathrm{b}1} = \begin{pmatrix}
    {0} & {1} & {0} & {0} & {0} & {0} & {0} \\
    {1} & {0} & {1} & {0} & {0} & {0} & {0} \\
    {0} & {1} & {0} & {1} & {0} & {0} & {0} \\
    {0} & {0} & {1} & {0} & {1} & {0} & {0} \\
    {0} & {0} & {0} & {1} & {0} & {1} & {0} \\
    {0} & {0} & {0} & {0} & {1} & {0} & {1} \\
    {0} & {0} & {0} & {0} & {0} & {1} & {0} \\
  \end{pmatrix},\,
  \bar{q}_0^{\mathrm{b}2} = \begin{pmatrix}
    {0} & {0} & {1} & {0} & {0} & {0} & {0} \\
    {0} & {0} & {0} & {1} & {0} & {0} & {0} \\
    {1} & {0} & {0} & {0} & {1} & {0} & {0} \\
    {0} & {1} & {0} & {0} & {0} & {1} & {0} \\
    {0} & {0} & {1} & {0} & {0} & {0} & {1} \\
    {0} & {0} & {0} & {1} & {0} & {0} & {0} \\
    {0} & {0} & {0} & {0} & {1} & {0} & {0} \\
  \end{pmatrix},\cdots,\\
  \bar{q}_0^{\mathrm{c}\prime} &= \begin{pmatrix}
    {0} & {0} & {0} & {0} & {0} & {0} & {0} \\
    {0} & {0} & {0} & {0} & {0} & {0} & {0} \\
    {0} & {0} & {1} & {0} & {0} & {0} & {0} \\
    {0} & {0} & {0} & {0} & {0} & {0} & {0} \\
    {0} & {0} & {0} & {0} & {1} & {0} & {0} \\
    {0} & {0} & {0} & {0} & {0} & {0} & {0} \\
    {0} & {0} & {0} & {0} & {0} & {0} & {0} \\
  \end{pmatrix},\,
  \bar{q}_0^{\mathrm{b}2\prime} = \begin{pmatrix}
    {0} & {0} & {0} & {0} & {0} & {0} & {0} \\
    {0} & {0} & {0} & {0} & {0} & {0} & {0} \\
    {0} & {0} & {0} & {0} & {1} & {0} & {0} \\
    {0} & {0} & {0} & {0} & {0} & {0} & {0} \\
    {0} & {0} & {1} & {0} & {0} & {0} & {0} \\
    {0} & {0} & {0} & {0} & {0} & {0} & {0} \\
    {0} & {0} & {0} & {0} & {0} & {0} & {0} \\
  \end{pmatrix},\,
  \bar{t}_x^{\mathrm{b}2\prime} = \begin{pmatrix}
    {0} & {0} & {0} & {0} & {0} & {0} & {0} \\
    {0} & {0} & {0} & {0} & {0} & {0} & {0} \\
    {0} & {0} & {0} & {0} & {-i} & {0} & {0} \\
    {0} & {0} & {0} & {0} & {0} & {0} & {0} \\
    {0} & {0} & {i} & {0} & {0} & {0} & {0} \\
    {0} & {0} & {0} & {0} & {0} & {0} & {0} \\
    {0} & {0} & {0} & {0} & {0} & {0} & {0} \\
  \end{pmatrix},\\
  \bar{q}_x^{\mathrm{c}} &= \begin{pmatrix}
    {-3} & {0} & {0} & {0} & {0} & {0} & {0} \\
    {0} & {-2} & {0} & {0} & {0} & {0} & {0} \\
    {0} & {0} & {-1} & {0} & {0} & {0} & {0} \\
    {0} & {0} & {0} & {0} & {0} & {0} & {0} \\
    {0} & {0} & {0} & {0} & {1} & {0} & {0} \\
    {0} & {0} & {0} & {0} & {0} & {2} & {0} \\
    {0} & {0} & {0} & {0} & {0} & {0} & {3} \\
  \end{pmatrix},\,
  \bar{q}_y^{\mathrm{c}} = \begin{pmatrix}
    {3} & {0} & {0} & {0} & {0} & {0} & {0} \\
    {0} & {2} & {0} & {0} & {0} & {0} & {0} \\
    {0} & {0} & {1} & {0} & {0} & {0} & {0} \\
    {0} & {0} & {0} & {0} & {0} & {0} & {0} \\
    {0} & {0} & {0} & {0} & {1} & {0} & {0} \\
    {0} & {0} & {0} & {0} & {0} & {2} & {0} \\
    {0} & {0} & {0} & {0} & {0} & {0} & {3} \\
  \end{pmatrix}.
\end{aligned}\end{align}

We display schematic pictures of the multipole bases 
$\bar{q}_0^{\mathrm{c}},\bar{q}_0^{\mathrm{b}1},\bar{q}_0^{\mathrm{b}2\prime},\bar{t}_x^{\mathrm{b}2\prime},\bar{q}_x^{\mathrm{c}},\bar{q}_y^{\mathrm{c}}$, and $\bar{\sigma}_{\nu}\,(\nu=x,y,z)$ in Fig.~\ref{fig:MpB}.

\begin{figure}[b]
  \centering
  \includegraphics{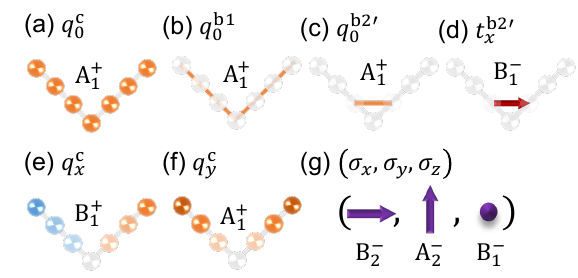}
  \caption{\label{fig:MpB}
  Schematic pictures of the multipole basis for the cluster, bond, and spin for the $7$-site basis,
  together with the corresponding irreducible representations under the point group $C_{\mathrm{2v}}$.
  The superscript $\pm$ of the irreducible representation denotes the parity with respect to the time-reversal operation.
  }
\end{figure}

\newpage % arXiv
\section{Multipole basis representation of the ME tensor} \label{sec:app:MpBME} %--------------------------------------------------------------------------------------------
We present the expressions for the ME tensor decomposed in the multipole basis, corresponding to Figs.~\ref{fig:MEprocess} and \ref{fig:ferromag}(b).
By separating the $\theta$ dependence of each operator in the definition of $\alpha^{z;x}_{\mathrm{eff}}$ in Eq.~(\ref{eq:effsus}), we obtain the following:
\begin{align}
  \begin{aligned}
    \alpha^{z;x}_{\mathrm{eff}}(i\omega_m)
    &=- \frac{k_{\mathrm{B}}T}{V} \sum_{i\epsilon_n}
    \Tr \left[ 
      \left(\bar{M}^{z(\pi)}_{\mathrm{eff}} + \varDelta\bar{M}^{z(\theta)}_{\mathrm{eff}} \right)
      \left(\bar{\mathcal{G}}^{+} + \bar{\mathcal{G}}^{+}\bar{\mathcal{T}}^{(\theta)+}\bar{\mathcal{G}}^{+} \right)
      \bar{P}^{x}
      \left(\bar{\mathcal{G}} + \bar{\mathcal{G}}\bar{\mathcal{T}}^{(\theta)}\bar{\mathcal{G}} \right)
    \right]\\
    &=- \frac{k_{\mathrm{B}}T}{V} \sum_{i\epsilon_n}
    \Tr \Bigl[
      \left(
        M_{q_0^{\mathrm{c}}\sigma_z} (\bar{q}_0^{\mathrm{c}}\bar{\sigma}_z) + M_{q_0^{\mathrm{b}}\sigma_z} (\bar{q}_0^{\mathrm{b}2}\bar{\sigma}_z) + \varDelta M_{q_0^{\mathrm{b}}\sigma_z} (\bar{q}_0^{\mathrm{b}2\prime}\bar{\sigma}_z) + \varDelta M_{t_x^{\mathrm{b}}\sigma_0}(\bar{t}_x^{\mathrm{b}2\prime}\bar{\sigma}_0)
      \right) \\
      & \qquad \times \left( 
        \bar{\mathcal{G}}^{+} + \bar{\mathcal{G}}^{+}
        \bigl[
        \mathcal{T}_{q_0^{\mathrm{c}}\sigma_0}^{+} (\bar{q}_0^{\mathrm{c}\prime}\bar{\sigma}_0) + \mathcal{T}_{q_0^{\mathrm{b}}\sigma_0}^{+} (\bar{q}_0^{\mathrm{b}2\prime}\bar{\sigma}_0) + \mathcal{T}_{t_x^{\mathrm{b}}\sigma_0}^{+} (\bar{t}_x^{\mathrm{b}2\prime}\bar{\sigma}_z)
        \bigr]
        \bar{\mathcal{G}}^{+} 
      \right)\\
      & \qquad \times \left(
        P_{q_x^{\mathrm{c}}\sigma_0} (\bar{q}_x^{\mathrm{c}}\bar{\sigma}_0)
      \right) \times
      \left( 
        \bar{\mathcal{G}} + \bar{\mathcal{G}}
        \bigl[
        \mathcal{T}_{q_0^{\mathrm{c}}\sigma_0} (\bar{q}_0^{\mathrm{c}\prime}\bar{\sigma}_0) + \mathcal{T}_{q_0^{\mathrm{b}}\sigma_0} (\bar{q}_0^{\mathrm{b}2\prime}\bar{\sigma}_0) + \mathcal{T}_{t_x^{\mathrm{b}}\sigma_0} (\bar{t}_x^{\mathrm{b}2\prime}\bar{\sigma}_z)
        \bigr]\bar{\mathcal{G}} 
      \right)
    \Bigr]\\
    &=\alpha^{z;x}_{\mathrm{i}}+\alpha^{z;x}_{\mathrm{ii}}+\alpha^{z;x}_{\mathrm{iii}}+\alpha^{z;x}_{\mathrm{iv}}
    +O(\Delta_p^{-4}),
  \end{aligned}
\end{align}
where $\bar{\mathcal{G}}$ stands for the simple 1D chain Green's function with $\theta=\pi$, the superscript ``$+$'' on the Green's function and the $T$-matrix indicates that their frequency arguments are $i\epsilon_n + i\omega_m$ [e.g., $ \bar{\mathcal{G}}^{+}=\bar{\mathcal{G}}(i\epsilon_n + i\omega_m)$].
$\mathcal{T}_{q_0^{\mathrm{c}}\sigma_0}$, $\mathcal{T}_{q_0^{\mathrm{b}}\sigma_0}$, $\mathcal{T}_{t_x^{\mathrm{b}}\sigma_z}$ $\varDelta M_{q_0^{\mathrm{b}}\sigma_z}$, and $\varDelta M_{t_x^{\mathrm{b}}\sigma_0}$ are the components of the $T$-matrix and the effective magnetization defined in Eqs.~(\ref{eq:Tcomp}) and (\ref{eq:dMcomp}).
$M_{q_0^{\mathrm{c}}\sigma_z}$, $M_{q_0^{\mathrm{b}}\sigma_z}$, $P_{q_x^{\mathrm{c}} \sigma_0}$, and $P_{q_y^{\mathrm{c}} \sigma_0}$ denote the $\theta=\pi$ components of the effective magnetization and the electric polarization operators along the $x$ and $y$ directions, respectively.
Their expressions are given by
\begin{align}\begin{aligned}
  M_{q_0^{\mathrm{c}}\sigma_z}  = \frac{a}{2V} \Tr[\,(\bar{q}_0^{\mathrm{c}}\bar{\sigma}_z)\,\bar{M}^{z(\pi)}_{\mathrm{eff}}\,] &,\quad M_{q_0^{\mathrm{b}}\sigma_z} = \frac{a}{2V} \Tr[\,(\bar{q}_0^{\mathrm{b}2}\bar{\sigma}_z)\,\bar{M}^{z(\pi)}_{\mathrm{eff}}\,]\\
  P_{q_x^{\mathrm{c}} \sigma_0} = a \sin{\frac{\theta}{2}} &,\quad P_{q_y^{\mathrm{c}} \sigma_0} = a \cos{\frac{\theta}{2}}.
\end{aligned}\end{align}
Eq.~(\ref{eq:METxSepa}) of $\alpha^{z;y}_{\mathrm{eff}}$ can also be obtained through a similar expansion.

With the condition $\Tr[\ast] \ne 0$ in mind, and using Eq.~(\ref{eq:selection}),
$\alpha^{z;x}_{\mathrm{i}}$, $\alpha^{z;x}_{\mathrm{ii}}$, $\alpha^{z;x}_{\mathrm{iii}}$, $\alpha^{z;x}_{\mathrm{iv}}$, and $\alpha_{\mathrm{Z}}^{z;y}$ are given as follows:
\begin{subequations}\begin{align}
  %% Type 1
  &\begin{aligned}
    \alpha^{z;x}_{\mathrm{i}}
    &=-\frac{k_{\mathrm{B}}T}{V} \sum_{i\epsilon_n}
    \Tr \left[ \, 
      (\bar{t}_x^{\mathrm{b}2\prime}\bar{\sigma}_0) \, 
      \bar{\mathcal{G}}^{+} \, 
      (\bar{q}_x^{\mathrm{c}}\bar{\sigma}_0)  \, 
      \bar{\mathcal{G}} \, 
    \right] \varDelta M_{t_x^{\mathrm{b}}\sigma_0} P_{q_x^{\mathrm{c}} \sigma_0},%\\
    %&\propto \Delta_p^{-2} \sin{\theta} \sin{\frac{\theta}{2}},
  \end{aligned}\\
  %% Type 2
  &\begin{aligned}
    \alpha^{z;x}_{\mathrm{ii}}&=
    -\frac{k_{\mathrm{B}}T}{V} \sum_{i\epsilon_n}
    \Tr \left[ \, 
      (\bar{t}_x^{\mathrm{b}2\prime}\bar{\sigma}_0) \, 
      \bar{\mathcal{G}}^{+} \, 
      (\bar{q}_x^{\mathrm{c}}\bar{\sigma}_0)  \, 
      \bar{\mathcal{G}} \, 
      (\bar{q}_0^{\mathrm{b}2\prime}\bar{\sigma}_0) \, 
      \bar{\mathcal{G}} \, 
    \right] \varDelta M_{t_x^{\mathrm{b}}\sigma_0} P_{q_x^{\mathrm{c}} \sigma_0} \mathcal{T}_{q_0^{\mathrm{b}}\sigma_0} \\
    &\qquad -\frac{k_{\mathrm{B}}T}{V} \sum_{i\epsilon_n}
    \Tr \left[ \, 
      (\bar{t}_x^{\mathrm{b}2\prime}\bar{\sigma}_0) \, 
      \bar{\mathcal{G}}^{+} \,
      (\bar{q}_0^{\mathrm{b}2\prime}\bar{\sigma}_0) \, 
      \bar{\mathcal{G}}^{+} \,
      (\bar{q}_x^{\mathrm{c}}\bar{\sigma}_0)  \, 
      \bar{\mathcal{G}}  \, 
    \right] \varDelta M_{t_x^{\mathrm{b}}\sigma_0} \mathcal{T}_{q_0^{\mathrm{b}}\sigma_0}^{+} P_{q_x^{\mathrm{c}} \sigma_0} ,%\\
    %& \sim \Delta_p^{-3} (1+\cos{\theta})  \sin{\theta} \sin{\frac{\theta}{2}},
  \end{aligned}\\
  %% Type 3
  &\begin{aligned}
    \alpha^{z;x}_{\mathrm{iii}}
    &=-\frac{k_{\mathrm{B}}T}{V} \sum_{i\epsilon_n}
    \Tr \left[ \, 
      (\bar{q}_0^{\mathrm{c}}\bar{\sigma}_z) \, 
      \bar{\mathcal{G}}^{+} \, 
      (\bar{q}_x^{\mathrm{c}}\bar{\sigma}_0)  \, 
      \bar{\mathcal{G}} \, 
      (\bar{t}_x^{\mathrm{b}2\prime}\bar{\sigma}_z) \, 
      \bar{\mathcal{G}} \, 
    \right] M_{q_0^{\mathrm{c}}\sigma_z} P_{q_x^{\mathrm{c}} \sigma_0} \mathcal{T}_{t_x^{\mathrm{b}}\sigma_z} \\
    &\qquad -\frac{k_{\mathrm{B}}T}{V} \sum_{i\epsilon_n}        
    \Tr \left[ \, 
      (\bar{q}_0^{\mathrm{c}}\bar{\sigma}_z) \, 
      \bar{\mathcal{G}}^{+} \,
      (\bar{t}_x^{\mathrm{b}2\prime}\bar{\sigma}_z) \, 
      \bar{\mathcal{G}}^{+} \,
      (\bar{q}_x^{\mathrm{c}}\bar{\sigma}_0)  \, 
      \bar{\mathcal{G}}  \, 
    \right] M_{q_0^{\mathrm{c}}\sigma_z} \mathcal{T}_{t_x^{\mathrm{b}}\sigma_z}^{+} P_{q_x^{\mathrm{c}} \sigma_0},%\\
    %& \sim \Delta_p^{-2} \sin{\theta} \sin{\frac{\theta}{2}},
  \end{aligned}\\
  %% Type 4
  &\begin{aligned}
    \alpha^{z;x}_{\mathrm{iv}}
    &=-\frac{k_{\mathrm{B}}T}{V} \sum_{i\epsilon_n}
    \Tr \left[ \, 
      (\bar{q}_0^{\mathrm{c}}\bar{\sigma}_0) \, 
      \bar{\mathcal{G}}^{+} \,
      (\bar{q}_0^{\mathrm{b}2\prime}\bar{\sigma}_z) \, 
      \bar{\mathcal{G}}^{+} \,
      (\bar{q}_x^{\mathrm{c}}\bar{\sigma}_0)  \, 
      \bar{\mathcal{G}} \, 
      (\bar{t}_x^{\mathrm{b}2\prime}\bar{\sigma}_z) \, 
      \bar{\mathcal{G}} \, 
    \right] M_{q_0^{\mathrm{c}}\sigma_z} \mathcal{T}_{q_0^{\mathrm{b}}\sigma_0}^{+} P_{q_x^{\mathrm{c}} \sigma_0} \mathcal{T}_{t_x^{\mathrm{b}}\sigma_z} \\
    &\qquad -\frac{k_{\mathrm{B}}T}{V} \sum_{i\epsilon_n}
    \Tr \left[ \, 
      (\bar{q}_0^{\mathrm{c}}\bar{\sigma}_0) \, 
      \bar{\mathcal{G}}^{+} \,
      (\bar{t}_x^{\mathrm{b}2\prime}\bar{\sigma}_z) \, 
      \bar{\mathcal{G}}^{+} \,
      (\bar{q}_x^{\mathrm{c}}\bar{\sigma}_0)  \, 
      \bar{\mathcal{G}} \, 
      (\bar{q}_0^{\mathrm{b}2\prime}\bar{\sigma}_z) \, 
      \bar{\mathcal{G}} \, 
    \right] M_{q_0^{\mathrm{c}}\sigma_z} \mathcal{T}_{t_x^{\mathrm{b}}\sigma_z}^{+} P_{q_x^{\mathrm{c}} \sigma_0} \mathcal{T}_{q_0^{\mathrm{b}}\sigma_0},%\\
    %& \sim \Delta_p^{-3} (1+\cos{\theta})  \sin{\theta} \sin{\frac{\theta}{2}},
  \end{aligned}\\
  %% Type 0
  &\begin{aligned}
    \alpha_{\mathrm{Z}}^{z;y}
    &=-\frac{k_{\mathrm{B}}T}{V} \sum_{i\epsilon_n}
    \Tr \left[ \, 
      (\bar{q}_0^{\mathrm{c}}\bar{\sigma}_0) \, 
      \bar{\mathcal{G}}^{+} \, 
      (\bar{q}_y^{\mathrm{c}}\bar{\sigma}_0)  \, 
      \bar{\mathcal{G}} \,
    \right] M_{q_0^{\mathrm{c}}\sigma_z} P_{q_y^{\mathrm{c}} \sigma_0}.
    %&\propto \Delta_p^{0} \cos{\frac{\theta}{2}},
  \end{aligned}
\end{align}\end{subequations}
From these equations, the $\theta$ and $\Delta_{p}$ dependence of the ME tensor in Eqs.~(\ref{eq:MEcontriddep}) and (\ref{eq:MEzeeddep}) in the main text is obtained.

\newpage % arXiv
\end{widetext}

%- bib  -------------------------------------------------------------------------------
\nocite{apsrev42Control}
\bibliographystyle{apsrev4-2}
\bibliography{KunojiME_bib.bib} % arXiv

@CONTROL{apsrev42Control,author="08",editor="1",pages="0",title="0",year="1"}

@PREAMBLE{
 "\providecommand{\noopsort}[1]{}" 
}

@article{cai2007optical,
  title={Optical cloaking with metamaterials},
  author={Cai, Wenshan and Chettiar, Uday K and Kildishev, Alexander V and Shalaev, Vladimir M},
  journal={Nat. Photon.},
  volume={1},
  number={4},
  pages={224--227},
  year={2007},
  publisher={Nature Publishing Group UK London},
  doi={https://doi.org/10.1038/nphoton.2007.28},
  url={https://www.nature.com/articles/nphoton.2007.28}
}

@article{soukoulis2011past,
  title={Past achievements and future challenges in the development of three-dimensional photonic metamaterials},
  author={Soukoulis, Costas M and Wegener, Martin},
  journal={Nat. Photon.},
  volume={5},
  number={9},
  pages={523--530},
  year={2011},
  publisher={Nature Publishing Group UK London},
  doi={https://doi.org/10.1038/nphoton.2011.154},
  url={https://www.nature.com/articles/nphoton.2011.154},
}

@article{zheludev2012metamaterials,
  title={From metamaterials to metadevices},
  author={Zheludev, Nikolay I and Kivshar, Yuri S},
  journal={Nat. Mater.},
  volume={11},
  number={11},
  pages={917--924},
  year={2012},
  publisher={Nature Publishing Group UK London},
  doi={https://doi.org/10.1038/nmat3431},
  url={https://www.nature.com/articles/nmat3431}
}

@article{https://doi.org/10.1002/smtd.201600064,
  author = {Hsiao, Hui-Hsin and Chu, Cheng Hung and Tsai, Din Ping},
  title = {Fundamentals and Applications of Metasurfaces},
  journal = {Small Methods},
  volume = {1},
  number = {4},
  pages = {1600064},
  doi = {https://doi.org/10.1002/smtd.201600064},
  url = {https://onlinelibrary.wiley.com/doi/abs/10.1002/smtd.201600064},
  year = {2017}
}

@article{PhysRevLett.95.227401,
  title = {Giant Optical Activity in Quasi-Two-Dimensional Planar Nanostructures},
  author = {Kuwata-Gonokami, Makoto and Saito, Nobuyoshi and Ino, Yusuke and Kauranen, Martti and Jefimovs, Konstantins and Vallius, Tuomas and Turunen, Jari and Svirko, Yuri},
  journal = {Phys. Rev. Lett.},
  volume = {95},
  issue = {22},
  pages = {227401},
  numpages = {4},
  year = {2005},
  month = {Nov},
  publisher = {American Physical Society},
  doi = {10.1103/PhysRevLett.95.227401},
  url = {https://link.aps.org/doi/10.1103/PhysRevLett.95.227401}
}

@article{soukoulis2010optical,
  title={Optical metamaterials—more bulky and less lossy},
  author={Soukoulis, Costas M and Wegener, Martin},
  journal={Science},
  volume={330},
  number={6011},
  pages={1633--1634},
  year={2010},
  publisher={American Association for the Advancement of Science},
  doi={10.1126/science.1198858},
  url={https://www.science.org/doi/10.1126/science.1198858},
}

@article{yu2011light,
  title={Light propagation with phase discontinuities: generalized laws of reflection and refraction},
  author={Yu, Nanfang and Genevet, Patrice and Kats, Mikhail A and Aieta, Francesco and Tetienne, Jean-Philippe and Capasso, Federico and Gaburro, Zeno},
  journal={Science},
  volume={334},
  number={6054},
  pages={333--337},
  year={2011},
  publisher={American Association for the Advancement of Science},
  doi={10.1126/science.1210713},
  url={https://www.science.org/doi/10.1126/science.1210713}
}

@article{gartside2021reconfigurable,
  title={Reconfigurable magnonic mode-hybridisation and spectral control in a bicomponent artificial spin ice},
  author={Gartside, Jack C and Vanstone, Alex and Dion, Troy and Stenning, Kilian D and Arroo, Daan M and Kurebayashi, Hidekazu and Branford, Will R},
  journal={Nat. Commun.},
  volume={12},
  number={1},
  pages={2488},
  year={2021},
  publisher={Nature Publishing Group UK London},
  doi={10.1038/s41467-021-22723-x},
  url={https://doi.org/10.1038/s41467-021-22723-x}
}

@article{10.1063/5.0089235,
  author = {Sekine, Daiki and Sato, Yoshifumi and Matsubara, Masakazu},
  title = {Nonlinear optical detection of mesoscopic magnetic toroidal dipoles},
  journal = {Appl. Phys. Lett.},
  volume = {120},
  number = {16},
  pages = {162905},
  year = {2022},
  month = {04},
  issn = {0003-6951},
  doi = {10.1063/5.0089235},
  url = {https://doi.org/10.1063/5.0089235},
}

@article{matsubara2022polarization,
  title={Polarization-controlled tunable directional spin-driven photocurrents in a magnetic metamaterial with threefold rotational symmetry},
  author={Matsubara, Masakazu and Kobayashi, Takatsugu and Watanabe, Hikaru and Yanase, Youichi and Iwata, Satoshi and Kato, Takeshi},
  journal={Nat. Commun.},
  volume={13},
  number={1},
  pages={6708},
  year={2022},
  publisher={Nature Publishing Group UK London},
  doi={10.1038/s41467-022-34374-7},
  url={https://www.nature.com/articles/s41467-022-34374-7}
}

@article{Cavannaetal2025,
  url = {https://doi.org/10.1515/nanoph-2025-0514},
  title = {Scaling-dependent tunability of spin-driven photocurrents in magnetic metamaterials},
  title = {},
  author = {Gabriele Cavanna and Hidehisa Taketani and Hikaru Watanabe and Da Pan and Anna Honda and Daiki Oshima and Takeshi Kato and Masakazu Matsubara},
  pages = {5611--5619},
  volume = {14},
  number = {27},
  journal = {Nanophotonics},
  doi  = {10.1515/nanoph-2025-0514},
  year = {2025},
}

@article{PhysRevB.107.155419,
  title = {Terahertz spin ratchet effect in magnetic metamaterials},
  author = {Hild, M. and Golub, L. E. and Fuhrmann, A. and Otteneder, M. and Kronseder, M. and Matsubara, M. and Kobayashi, T. and Oshima, D. and Honda, A. and Kato, T. and Wunderlich, J. and Back, C. and Ganichev, S. D.},
  journal = {Phys. Rev. B},
  volume = {107},
  issue = {15},
  pages = {155419},
  numpages = {15},
  year = {2023},
  month = {Apr},
  publisher = {American Physical Society},
  doi = {10.1103/PhysRevB.107.155419},
  url = {https://link.aps.org/doi/10.1103/PhysRevB.107.155419}
}

@ARTICLE{7109859,
  author = {Gao, Xi and Han, Xu and Cao, Wei-Ping and Li, Hai Ou and Ma, Hui Feng and Cui, Tie Jun},
  journal= {IEEE Trans. Antennas Propag.}, 
  title  = {Ultrawideband and High-Efficiency Linear Polarization Converter Based on Double V-Shaped Metasurface}, 
  year   = {2015},
  volume = {63},
  number = {8},
  pages  = {3522-3530},
  doi = {10.1109/TAP.2015.2434392}
}

@article{xie2019anomalous,
  title={Anomalous refraction and reflection characteristics of bend V-shaped antenna metasurfaces},
  author={Xie, Yanqiang and Yang, Chang and Wang, Yun and Shen, Yun and Deng, Xiaohua and Zhou, Binbin and Cao, Juncheng},
  journal={Scientific reports},
  volume={9},
  number={1},
  pages={6700},
  year={2019},
  publisher={Nature Publishing Group UK London},
  doi={10.1038/s41598-019-43138-1}
}

@article{hu2021review,
  title={A review on metasurface: from principle to smart metadevices},
  author={Hu, Jie and Bandyopadhyay, Sankhyabrata and Liu, Yu-hui and Shao, Li-yang},
  journal={Frontiers in Physics},
  volume={8},
  pages={586087},
  year={2021},
  publisher={Frontiers Media SA},
  doi={10.3389/fphy.2020.586087}
}

@article{Babynina2016,
  author = {Babynina, Anastasia and Fedoruk, Michael and K{\"u}hler, Paul and Meledin, Alexander and D{\"o}blinger, Markus and Lohm{\"u}ller, Theobald},
  title = {Bending Gold Nanorods with Light},
  journal = {Nano Letters},
  volume = {16},
  number = {10},
  pages = {6485-6490},
  year = {2016},
  month = {10},
  issn = {1530-6984},
  doi = {10.1021/acs.nanolett.6b03029},
  url = {https://doi.org/10.1021/acs.nanolett.6b03029},
}

@article{Billot15,
  author = {L. Billot and M. Mortier and L. Aigouy},
  journal = {Appl. Opt.},
  number = {31},
  pages = {9326--9330},
  publisher = {Optica Publishing Group},
  title = {Near-field observation of surface plasmon polaritons launched by V-shaped nanorods on a gold surface},
  volume = {54},
  month = {Nov},
  year = {2015},
  url = {https://opg.optica.org/ao/abstract.cfm?URI=ao-54-31-9326},
  doi = {10.1364/AO.54.009326},
}

@article{cai2010atomically,
  title  = {Atomically precise bottom-up fabrication of graphene nanoribbons},
  author = {
    Cai, Jinming and 
    Ruffieux, Pascal and 
    Jaafar, Rached and 
    Bieri, Marco and 
    Braun, Thomas and 
    Blankenburg, Stephan and 
    Muoth, Matthias and 
    Seitsonen, Ari P and 
    Saleh, Moussa and 
    Feng, Xinliang and 
    Müllen, Klaus and 
    Fasel, Roman
  },
  journal  = {Nature},
  volume   = {466},
  number   = {7305},
  pages    = {470--473},
  year     = {2010},
  publisher= {Nature Publishing Group UK London},
  url      = {https://doi.org/10.1038/nature09211}
}

@article{cao2018correlated,
  title={Correlated insulator behaviour at half-filling in magic-angle graphene superlattices},
  author={
    Cao, Yuan and 
    Fatemi, Valla and 
    Demir, Ahmet and 
    Fang, Shiang and 
    Tomarken, Spencer L and 
    Luo, Jason Y and 
    Sanchez-Yamagishi, Javier D and 
    Watanabe, Kenji and 
    Taniguchi, Takashi and 
    Kaxiras, Efthimios and
    Ashoori, Ray C. and 
    Jarillo-Herrero, Pablo
  },
  journal={Nature},
  volume={556},
  number={7699},
  pages={80--84},
  year={2018},
  publisher={Nature Publishing Group UK London},
  url ={https://doi.org/10.1038/nature26154}
}

@article{Houtsma2021,
  author = {Houtsma, R. S. Koen and de la Rie, Joris and Stöhr, Meike},
  title = {Atomically precise graphene nanoribbons: interplay of structural and electronic properties},
  journal = {Chemical Society Reviews},
  volume = {50},
  number = {11},
  pages = {6541-6568},
  year = {2021},
  month = {06},
  issn = {0306-0012},
  doi = {10.1039/d0cs01541e},
  url = {https://doi.org/10.1039/d0cs01541e},
}

@article{Link1999,
  author = {Link, S. and Mohamed, M. B. and El-Sayed, M. A.},
  title = {Simulation of the Optical Absorption Spectra of Gold Nanorods as a Function of Their Aspect Ratio and the Effect of the Medium Dielectric Constant},
  journal = {The Journal of Physical Chemistry B},
  volume = {103},
  number = {16},
  pages = {3073-3077},
  year = {1999},
  month = {04},
  issn = {1520-6106},
  doi = {10.1021/jp990183f},
  url = {https://doi.org/10.1021/jp990183f},
}

@article{Jain2006,
  author = {Jain, Prashant K. and Lee, Kyeong Seok and El-Sayed, Ivan H. and El-Sayed, Mostafa A.},
  title = {Calculated Absorption and Scattering Properties of Gold Nanoparticles of Different Size, Shape, and Composition: Applications in Biological Imaging and Biomedicine},
  journal = {The Journal of Physical Chemistry B},
  volume = {110},
  number = {14},
  pages = {7238-7248},
  year = {2006},
  month = {04},
  issn = {1520-6106},
  doi = {10.1021/jp057170o},
  url = {https://doi.org/10.1021/jp057170o},
}

@article{PhysRevB.98.245129,
  title = {Group-theoretical classification of multipole order: Emergent responses and candidate materials},
  author = {Watanabe, Hikaru and Yanase, Youichi},
  journal = {Phys. Rev. B},
  volume = {98},
  issue = {24},
  pages = {245129},
  numpages = {24},
  year = {2018},
  month = {Dec},
  publisher = {American Physical Society},
  doi = {10.1103/PhysRevB.98.245129},
  url = {https://link.aps.org/doi/10.1103/PhysRevB.98.245129}
}

@article{PhysRevB.102.144441,
  title = {Bottom-up design of spin-split and reshaped electronic band structures in antiferromagnets without spin-orbit coupling: Procedure on the basis of augmented multipoles},
  author = {Hayami, Satoru and Yanagi, Yuki and Kusunose, Hiroaki},
  journal = {Phys. Rev. B},
  volume = {102},
  issue = {14},
  pages = {144441},
  numpages = {24},
  year = {2020},
  month = {Oct},
  publisher = {American Physical Society},
  doi = {10.1103/PhysRevB.102.144441},
  url = {https://link.aps.org/doi/10.1103/PhysRevB.102.144441}
}

@article{kusunose2022generalization,
  title={Generalization of microscopic multipoles and cross-correlated phenomena by their orderings},
  author={Kusunose, Hiroaki and Hayami, Satoru},
  journal={J. Phys.: Condens. Matter},
  volume={34},
  number={46},
  pages={464002},
  year={2022},
  publisher={IOP Publishing},
  doi={10.1088/1361-648X/ac9209},
  url={https://iopscience.iop.org/article/10.1088/1361-648X/ac9209/}
}

@article{hayami2024unified,
  title={Unified description of electronic orderings and cross correlations by complete multipole representation},
  author={Hayami, Satoru and Kusunose, Hiroaki},
  journal={J. Phys. Soc. Jpn.},
  volume={93},
  number={7},
  pages={072001},
  year={2024},
  publisher={The Physical Society of Japan},
  doi={https://doi.org/10.7566/JPSJ.93.072001}
}

@article{doi:10.7566/JPSJ.87.033709,
  author = {Hayami ,Satoru and Kusunose ,Hiroaki},
  title = {Microscopic Description of Electric and Magnetic Toroidal Multipoles in Hybrid Orbitals},
  journal = {J. Phys. Soc. Jpn.},
  volume = {87},
  number = {3},
  pages = {033709},
  year = {2018},
  doi = {10.7566/JPSJ.87.033709},
  URL = { https://doi.org/10.7566/JPSJ.87.033709}
}

@article{Hayami_PhysRevLett.122.147602,
  title = {{Electric Toroidal Quadrupoles in the Spin-Orbit-Coupled Metal {${\mathrm{Cd}}_{2}{\mathrm{Re}}_{2}{\mathrm{O}}_{7}$}}},
  author = {Hayami, Satoru and Yanagi, Yuki and Kusunose, Hiroaki and Motome, Yukitoshi},
  journal = {Phys. Rev. Lett.},
  volume = {122},
  issue = {14},
  pages = {147602},
  numpages = {6},
  year = {2019},
  month = {Apr},
  publisher = {American Physical Society},
  doi = {10.1103/PhysRevLett.122.147602},
  url = {https://link.aps.org/doi/10.1103/PhysRevLett.122.147602}
}

@article{doi:10.7566/JPSJ.89.104704,
  author = {Kusunose ,Hiroaki and Oiwa ,Rikuto and Hayami ,Satoru},
  title = {Complete Multipole Basis Set for Single-Centered Electron Systems},
  journal = {J. Phys. Soc. Jpn.},
  volume = {89},
  number = {10},
  pages = {104704},
  year = {2020},
  doi = {10.7566/JPSJ.89.104704},
  URL = { https://doi.org/10.7566/JPSJ.89.104704}
}

@article{Kusunose_PhysRevB.107.195118,
  title = {Symmetry-adapted modeling for molecules and crystals},
  author = {Kusunose, Hiroaki and Oiwa, Rikuto and Hayami, Satoru},
  journal = {Phys. Rev. B},
  volume = {107},
  issue = {19},
  pages = {195118},
  numpages = {14},
  year = {2023},
  month = {May},
  publisher = {American Physical Society},
  doi = {10.1103/PhysRevB.107.195118},
  url = {https://link.aps.org/doi/10.1103/PhysRevB.107.195118}
}

@article{2s5q-p42x,
  title = {Symmetry-adapted closest Wannier modeling based on complete multipole basis set},
  author = {Oiwa, Rikuto and Inda, Akane and Hayami, Satoru and Nomoto, Takuya and Arita, Ryotaro and Kusunose, Hiroaki},
  journal = {Phys. Rev. B},
  volume = {112},
  issue = {3},
  pages = {035116},
  numpages = {21},
  year = {2025},
  month = {Jul},
  publisher = {American Physical Society},
  doi = {10.1103/2s5q-p42x},
  url = {https://link.aps.org/doi/10.1103/2s5q-p42x}
}

@article{PhysRevB.98.165110,
  title = {Classification of atomic-scale multipoles under crystallographic point groups and application to linear response tensors},
  author = {Hayami, Satoru and Yatsushiro, Megumi and Yanagi, Yuki and Kusunose, Hiroaki},
  journal = {Phys. Rev. B},
  volume = {98},
  issue = {16},
  pages = {165110},
  numpages = {35},
  year = {2018},
  month = {Oct},
  publisher = {American Physical Society},
  doi = {10.1103/PhysRevB.98.165110},
  url = {https://link.aps.org/doi/10.1103/PhysRevB.98.165110}
}

@article{PhysRevB.104.054412,
  title = {Multipole classification in {$122$} magnetic point groups for unified understanding of multiferroic responses and transport phenomena},
  author = {Yatsushiro, Megumi and Kusunose, Hiroaki and Hayami, Satoru},
  journal = {Phys. Rev. B},
  volume = {104},
  issue = {5},
  pages = {054412},
  numpages = {41},
  year = {2021},
  month = {Aug},
  publisher = {American Physical Society},
  doi = {10.1103/PhysRevB.104.054412},
  url = {https://link.aps.org/doi/10.1103/PhysRevB.104.054412}
}

@article{doi:10.7566/JPSJ.91.014701,
  author = {Oiwa ,Rikuto and Kusunose ,Hiroaki},
  title = {Systematic Analysis Method for Nonlinear Response Tensors},
  journal = {J. Phys. Soc. Jpn.},
  volume = {91},
  number = {1},
  pages = {014701},
  year = {2022},
  doi = {10.7566/JPSJ.91.014701},
  URL = {https://doi.org/10.7566/JPSJ.91.014701}
}

@article{EDELSTEIN1990233,
  title = {Spin polarization of conduction electrons induced by electric current in two-dimensional asymmetric electron systems},
  journal = {Solid State Commun.},
  volume = {73},
  number = {3},
  pages = {233-235},
  year = {1990},
  issn = {0038-1098},
  doi = {https://doi.org/10.1016/0038-1098(90)90963-C},
  url = {https://www.sciencedirect.com/science/article/pii/003810989090963C},
  author = {V.M. Edelstein}
}

@article{furukawa2017observation,
  title={Observation of current-induced bulk magnetization in elemental tellurium},
  author={Furukawa, Tetsuya and Shimokawa, Yuri and Kobayashi, Kaya and Itou, Tetsuaki},
  journal={Nat. Commun.},
  volume={8},
  number={1},
  pages={954},
  year={2017},
  publisher={Nature Publishing Group},
  doi = {10.1038/s41467-017-01093-3},
}

@article{gonzalez2024non,
  title={Non-relativistic torque and Edelstein effect in non-collinear magnets},
  author={Gonz{\'a}lez-Hern{\'a}ndez, Rafael and Ritzinger, Philipp and V{\`y}born{\`y}, Karel and {\v{Z}}elezn{\`y}, Jakub and Manchon, Aur{\'e}lien},
  journal={Nat. Commun.},
  volume={15},
  number={1},
  pages={7663},
  year={2024},
  publisher={Nature Publishing Group UK London},
  doi={10.1038/s41467-024-51565-6},
}

@article{gkt2-x7mm,
  title = {Mechanically and electrically tunable Rashba-Edelstein effect in ferroelectric semiconductors, $\mathrm{CsGe}{X}_{3}$ $({X}=\mathrm{I}, \mathrm{Br}, \mathrm{Cl})$},
  author = {Popoola, Abduljelili and Kashikar, Ravi and Azmy, Ali and Spanopoulos, Ioannis and Jafari, Homayoun and S\l{}awi\ifmmode \acute{n}\else \'{n}\fi{}ska, Jagoda and Witanachchi, Sarath and Lisenkov, Sergey and Ponomareva, Inna},
  journal = {Phys. Rev. Mater.},
  volume = {9},
  issue = {8},
  pages = {084412},
  numpages = {8},
  year = {2025},
  month = {Aug},
  publisher = {American Physical Society},
  doi = {10.1103/gkt2-x7mm},
  url = {https://link.aps.org/doi/10.1103/gkt2-x7mm}
}

@article{PhysRevLett.115.216806,
  title = {Quantum Nonlinear Hall Effect Induced by Berry Curvature Dipole in Time-Reversal Invariant Materials},
  author = {Sodemann, Inti and Fu, Liang},
  journal = {Phys. Rev. Lett.},
  volume = {115},
  issue = {21},
  pages = {216806},
  numpages = {5},
  year = {2015},
  month = {Nov},
  publisher = {American Physical Society},
  doi = {10.1103/PhysRevLett.115.216806},
  url = {https://link.aps.org/doi/10.1103/PhysRevLett.115.216806}
}

@article{PhysRevB.95.235434,
  title = {Electronic and optical properties of the monolayer group-IV monochalcogenides $MX$ ($M=\mathrm{Ge},\phantom{\rule{4pt}{0ex}}\mathrm{Sn}$; $X=\mathrm{S},\phantom{\rule{4pt}{0ex}}\mathrm{Se},\phantom{\rule{4pt}{0ex}}\mathrm{Te}$)},
  author = {Xu, Lei and Yang, Ming and Wang, Shi Jie and Feng, Yuan Ping},
  journal = {Phys. Rev. B},
  volume = {95},
  issue = {23},
  pages = {235434},
  numpages = {9},
  year = {2017},
  month = {Jun},
  publisher = {American Physical Society},
  doi = {10.1103/PhysRevB.95.235434},
  url = {https://link.aps.org/doi/10.1103/PhysRevB.95.235434}
}

@article{ma2019observation,
  title={Observation of the nonlinear Hall effect under time-reversal-symmetric conditions},
  author={
    Ma, Qiong and 
    Xu, Su-Yang and 
    Shen, Huitao and 
    MacNeill, David and 
    Fatemi, Valla and 
    Chang, Tay-Rong and 
    Mier Valdivia, Andr{\'e}s M and 
    Wu, Sanfeng and 
    Du, Zongzheng and 
    Hsu, Chuang-Han and 
    Fang, Shiang and 
    Gibson, Quinn D. and
    Watanabe, Kenji and
    Taniguchi, Takashi and
    Cava, Robert J. and
    Kaxiras, Efthimios and
    Lu, Hai-Zhou and
    Lin, Hsin and
    Fu, Liang and
    Gedik, Nuh and
    Jarillo-Herrero, Pablo
  },
  journal={Nature (London)},
  volume={565},
  number={7739},
  pages={337--342},
  year={2019},
  publisher={Nature Publishing Group UK London},
  doi={10.1038/s41586-018-0807-6},
}

@article{wang2019ferroelectric,
  title={Ferroelectric nonlinear anomalous Hall effect in few-layer {$\mathrm{WTe}_2$}},
  author={Wang, Hua and Qian, Xiaofeng},
  journal={npj Computational Materials},
  volume={5},
  number={1},
  pages={119},
  year={2019},
  publisher={Nature Publishing Group UK London},
  doi={10.1038/s41524-019-0257-1}
}

@article{nakamura2017shift,
  title={Shift current photovoltaic effect in a ferroelectric charge-transfer complex},
  author={Nakamura, M and Horiuchi, S and Kagawa, F and Ogawa, N and Kurumaji, T and Tokura, Y and Kawasaki, M},
  journal={Nat. Commun.},
  volume={8},
  number={1},
  pages={281},
  year={2017},
  publisher={Nature Publishing Group UK London},
  doi={10.1038/s41467-017-00250-y},
}

@article{PhysRevB.96.241203,
  title = {Shift current in the ferroelectric semiconductor $\mathrm{SbSI}$},
  author = {Ogawa, N. and Sotome, M. and Kaneko, Y. and Ogino, M. and Tokura, Y.},
  journal = {Phys. Rev. B},
  volume = {96},
  issue = {24},
  pages = {241203},
  numpages = {4},
  year = {2017},
  month = {Dec},
  publisher = {American Physical Society},
  doi = {10.1103/PhysRevB.96.241203},
  url = {https://link.aps.org/doi/10.1103/PhysRevB.96.241203}
}

@article{shin2020dynamical,
  title={Dynamical amplification of electric polarization through nonlinear phononics in 2{D} $\mathrm{SnTe}$},
  author={Shin, Dongbin and Sato, Shunsuke A and H{\"u}bener, Hannes and De Giovannini, Umberto and Park, Noejung and Rubio, Angel},
  journal={npj Computational Materials},
  volume={6},
  number={1},
  pages={182},
  year={2020},
  publisher={Nature Publishing Group UK London},
  doi={10.1038/s41524-020-00449-6}
}

@article{okyay2022second,
  title={Second harmonic Hall responses of insulators as a probe of Berry curvature dipole},
  author={Okyay, Mahmut Sait and Sato, Shunsuke A and Kim, Kun Woo and Yan, Binghai and Jin, Hosub and Park, Noejung},
  journal={Commun. Phys.},
  volume={5},
  number={1},
  pages={303},
  year={2022},
  publisher={Nature Publishing Group UK London},
  doi={10.1038/s42005-022-01086-9},
}

@incollection{resta2007theory,
  title={Theory of polarization: a modern approach},
  author={Resta, Raffaele and Vanderbilt, David},
  booktitle={Physics of ferroelectrics: a modern perspective},
  pages={31--68},
  publisher={Springer},
  year={2007}
}

@article{cui2018two,
  title={Two-dimensional materials with piezoelectric and ferroelectric functionalities},
  author={Cui, Chaojie and Xue, Fei and Hu, Wei-Jin and Li, Lain-Jong},
  journal={npj 2D Materials and Applications},
  volume={2},
  number={1},
  pages={18},
  year={2018},
  publisher={Nature Publishing Group UK London},
  doi={10.1038/s41699-018-0063-5}
}

@article{10.1063/5.0251679,
    author = {Poplavko, Yuriy},
    title = {Piezoelectric intrinsic polarity modeling and determination},
    journal = {APL Materials},
    volume = {13},
    number = {6},
    pages = {061113},
    year = {2025},
    month = {06},
    issn = {2166-532X},
    doi = {10.1063/5.0251679},
    url = {https://doi.org/10.1063/5.0251679},
}

@article{DUBOVIK1990145,
  title = {Toroid moments in electrodynamics and solid-state physics},
  journal = {Phys. Rep.},
  volume = {187},
  number = {4},
  pages = {145-202},
  year = {1990},
  issn = {0370-1573},
  doi = {https://doi.org/10.1016/0370-1573(90)90042-Z},
  url = {https://www.sciencedirect.com/science/article/pii/037015739090042Z},
  author = {V.M. Dubovik and V.V. Tugushev}
}

@article{Spaldin_2008,
  doi = {10.1088/0953-8984/20/43/434203},
  url = {https://doi.org/10.1088/0953-8984/20/43/434203},
  year = {2008},
  month = {oct},
  publisher = {},
  volume = {20},
  number = {43},
  pages = {434203},
  author = {Spaldin, Nicola A and Fiebig, Manfred and Mostovoy, Maxim},
  title = {The toroidal moment in condensed-matter physics and its relation to the magnetoelectriceffect*},
  journal = {J. Phys.: Condensed Matter}
}

@article{kopaev2009toroidal,
  title={Toroidal ordering in crystals},
  author={Kopaev, Yurii Vasil'evich},
  journal={Physics-Uspekhi},
  volume={52},
  number={11},
  pages={1111--1125},
  year={2009},
  publisher={Turpion Ltd},
  doi = {10.3367/UFNe.0179.200911d.1175},
}

@article{tokura2018nonreciprocal,
  title={Nonreciprocal responses from non-centrosymmetric quantum materials},
  author={Tokura, Yoshinori and Nagaosa, Naoto},
  journal={Nat. Commun.},
  volume={9},
  number={1},
  pages={3740},
  year={2018},
  publisher={Nature Publishing Group UK London},
  doi={10.1038/s41467-018-05759-4}
}

@article{doi:10.7566/JPSJ.91.115001,
  author = {Isobe ,Hiroki and Nagaosa ,Naoto},
  title = {Toroidal Scattering and Nonreciprocal Transport by Magnetic Impurities},
  journal = {J. Phys. Soc. Jpn.},
  volume = {91},
  number = {11},
  pages = {115001},
  year = {2022},
  doi = {10.7566/JPSJ.91.115001},
  URL = {https://doi.org/10.7566/JPSJ.91.115001}
}

@article{doi:10.7566/JPSJ.94.083705,
  author = {Yamanaka ,Taisei and Ihara ,Yoshihiko and Hayami ,Satoru},
  title = {Nonlinear Nonreciprocal Electric Conductivity Driven by Magnetic Field Gradients},
  journal = {J. Phys. Soc. Jpn.},
  volume = {94},
  number = {8},
  pages = {083705},
  year = {2025},
  doi = {10.7566/JPSJ.94.083705},
  URL = {https://doi.org/10.7566/JPSJ.94.083705}
}

@article{Fiebig_2005,
  doi = {10.1088/0022-3727/38/8/R01},
  url = {https://doi.org/10.1088/0022-3727/38/8/R01},
  year = {2005},
  month = {apr},
  publisher = {},
  volume = {38},
  number = {8},
  pages = {R123},
  author = {Fiebig, Manfred},
  title = {Revival of the magnetoelectric effect},
  journal = {J. Phys. D: Appl. Phys.}
}

@article{hayami2016emergent,
  title={Emergent spin-valley-orbital physics by spontaneous parity breaking},
  author={Hayami, Satoru and Kusunose, Hiroaki and Motome, Yukitoshi},
  journal={J. Phys.: Condens. Matter},
  volume={28},
  number={39},
  pages={395601},
  year={2016},
  publisher={IOP Publishing},
  doi={10.1088/0953-8984/28/39/395601},
}

@article{doi:10.7566/JPSJ.87.033702,
  author = {Saito ,Hiraku and Uenishi ,Kenta and Miura ,Naoyuki and Tabata ,Chihiro and Hidaka ,Hiroyuki and Yanagisawa ,Tatsuya and Amitsuka ,Hiroshi},
  title = {Evidence of a New Current-Induced Magnetoelectric Effect in a Toroidal Magnetic Ordered State of UNi4B},
  journal = {J. Phys. Soc. Jpn.},
  volume = {87},
  number = {3},
  pages = {033702},
  year = {2018},
  doi = {10.7566/JPSJ.87.033702},
  URL = {https://doi.org/10.7566/JPSJ.87.033702}
}

@article{PhysRevB.97.134423,
  title = {Microscopic theory of spin toroidization in periodic crystals},
  author = {Gao, Yang and Vanderbilt, David and Xiao, Di},
  journal = {Phys. Rev. B},
  volume = {97},
  issue = {13},
  pages = {134423},
  numpages = {8},
  year = {2018},
  month = {Apr},
  publisher = {American Physical Society},
  doi = {10.1103/PhysRevB.97.134423},
  url = {https://link.aps.org/doi/10.1103/PhysRevB.97.134423}
}

@article{ding2021field,
  title={Field-tunable toroidal moment in a chiral-lattice magnet},
  author={
    Ding, Lei and 
    Xu, Xianghan and 
    Jeschke, Harald O and 
    Bai, Xiaojian and 
    Feng, Erxi and 
    Alemayehu, Admasu Solomon and 
    Kim, Jaewook and 
    Huang, Fei-Ting and 
    Zhang, Qiang and 
    Ding, Xiaxin and 
    Harrison, Neil and
    Zapf, Vivien and
    Khomskii, Daniel and
    Mazin, Igor I. and
    Cheong, Sang-Wook and
    Cao, Huibo
  },
  journal={Nat. Commun.},
  volume={12},
  number={1},
  pages={5339},
  year={2021},
  publisher={Nature Publishing Group UK London},
  doi={10.1038/s41467-021-25657-6},
}

@article{PhysRevB.105.155157,
  title = {Analysis of model-parameter dependences on the second-order nonlinear conductivity in $\mathcal{PT}$-symmetric collinear antiferromagnetic metals with magnetic toroidal moment on zigzag chains},
  author = {Yatsushiro, Megumi and Oiwa, Rikuto and Kusunose, Hiroaki and Hayami, Satoru},
  journal = {Phys. Rev. B},
  volume = {105},
  issue = {15},
  pages = {155157},
  numpages = {10},
  year = {2022},
  month = {Apr},
  publisher = {American Physical Society},
  doi = {10.1103/PhysRevB.105.155157},
  url = {https://link.aps.org/doi/10.1103/PhysRevB.105.155157}
}

@article{PhysRevB.111.L201112,
  title = {Finite-$q$ antiferrotoroidal and ferritoroidal order in a distorted kagome structure},
  author = {Kirikoshi, Akimitsu and Hayami, Satoru},
  journal = {Phys. Rev. B},
  volume = {111},
  issue = {20},
  pages = {L201112},
  numpages = {6},
  year = {2025},
  month = {May},
  publisher = {American Physical Society},
  doi = {10.1103/PhysRevB.111.L201112},
  url = {https://link.aps.org/doi/10.1103/PhysRevB.111.L201112}
}

@article{PhysRevLett.95.237402,
  title = {Optical Magnetoelectric Effect in Multiferroic Materials: Evidence for a Lorentz Force Acting on a Ray of Light},
  author = {Sawada, Kei and Nagaosa, Naoto},
  journal = {Phys. Rev. Lett.},
  volume = {95},
  issue = {23},
  pages = {237402},
  numpages = {4},
  year = {2005},
  month = {Dec},
  publisher = {American Physical Society},
  doi = {10.1103/PhysRevLett.95.237402},
  url = {https://link.aps.org/doi/10.1103/PhysRevLett.95.237402}
}

@article{doi:10.1143/JPSJ.81.023712,
  author = {Miyahara ,Shin and Furukawa ,Nobuo},
  title = {Nonreciprocal Directional Dichroism and Toroidalmagnons in Helical Magnets},
  journal = {J. Phys. Soc. Jpn.},
  volume = {81},
  number = {2},
  pages = {023712},
  year = {2012},
  doi = {10.1143/JPSJ.81.023712},
  URL = {https://doi.org/10.1143/JPSJ.81.023712}
}

@article{Hayami_doi:10.7566/JPSJ.85.053705,
author = {Satoru Hayami and  Hiroaki Kusunose and  Yukitoshi Motome},
title = {Asymmetric Magnon Excitation by Spontaneous Toroidal Ordering},
journal = {J. Phys. Soc. Jpn.},
volume = {85},
number = {5},
pages = {053705},
year = {2016},
doi = {10.7566/JPSJ.85.053705},
}

@article{PhysRevB.103.L180410,
  title = {Toroidal nonreciprocity of optical second harmonic generation},
  author = {Mund, J. and Yakovlev, D. R. and Poddubny, A. N. and Dubrovin, R. M. and Bayer, M. and Pisarev, R. V.},
  journal = {Phys. Rev. B},
  volume = {103},
  issue = {18},
  pages = {L180410},
  numpages = {6},
  year = {2021},
  month = {May},
  publisher = {American Physical Society},
  doi = {10.1103/PhysRevB.103.L180410},
  url = {https://link.aps.org/doi/10.1103/PhysRevB.103.L180410}
}

@article{Suzuki_PhysRevB.105.075201,
  title = {Tunneling spin current in systems with spin degeneracy},
  author = {Suzuki, Yuta},
  journal = {Phys. Rev. B},
  volume = {105},
  issue = {7},
  pages = {075201},
  numpages = {11},
  year = {2022},
  month = {Feb},
  publisher = {American Physical Society},
  doi = {10.1103/PhysRevB.105.075201},
  url = {https://link.aps.org/doi/10.1103/PhysRevB.105.075201}
}

@article{10.1063/1.1724312,
  author = {Löwdin, Per-Olov},
  title = {Studies in Perturbation Theory. IV. Solution of Eigenvalue Problem by Projection Operator Formalism},
  journal = {J. Math. Phys.},
  volume = {3},
  number = {5},
  pages = {969-982},
  year = {1962},
  month = {09},
  issn = {0022-2488},
  doi = {10.1063/1.1724312},
  url = {https://doi.org/10.1063/1.1724312},
}

@article{RevModPhys.36.1076,
  title = {Unified Theory of Nuclear Reactions},
  author = {Feshbach, Herman},
  journal = {Rev. Mod. Phys.},
  volume = {36},
  issue = {4},
  pages = {1076--1078},
  numpages = {0},
  year = {1964},
  month = {Oct},
  publisher = {American Physical Society},
  doi = {10.1103/RevModPhys.36.1076},
  url = {https://link.aps.org/doi/10.1103/RevModPhys.36.1076}
}

@article{LEINAAS197819,
  title = {Convergence properties of the Brillouin-Wigner type of perturbation expansion},
  journal = {Annals of Physics},
  volume = {111},
  number = {1},
  pages = {19-37},
  year = {1978},
  issn = {0003-4916},
  doi = {https://doi.org/10.1016/0003-4916(78)90222-1},
  url = {https://www.sciencedirect.com/science/article/pii/0003491678902221},
  author = {J.M Leinaas and T.T.S Kuo}
}

@article{10.1063/1.4904200,
    author = {Hatano, Naomichi and Ordonez, Gonzalo},
    title = {Time-reversal symmetric resolution of unity without background integrals in open quantum systems},
    journal = {J. Math. Phys.},
    volume = {55},
    number = {12},
    pages = {122106},
    year = {2014},
    month = {12},
    issn = {0022-2488},
    doi = {10.1063/1.4904200},
    url = {https://doi.org/10.1063/1.4904200},
}

@book{Altland_Simons_2010, 
  place={Cambridge}, 
  edition={2}, 
  title={Condensed Matter Field Theory}, 
  publisher={Cambridge University Press}, 
  author={Altland, Alexander and Simons, Ben D.}, 
  year={2010}
}

@article{PhysRevB.110.165111,
  title = {Impact of electron correlations on the nonlinear Edelstein effect},
  author = {\ifmmode \bar{O}\else \={O}\fi{}ik\'e, Jun and Peters, Robert},
  journal = {Phys. Rev. B},
  volume = {110},
  issue = {16},
  pages = {165111},
  numpages = {18},
  year = {2024},
  month = {Oct},
  publisher = {American Physical Society},
  doi = {10.1103/PhysRevB.110.165111},
  url = {https://link.aps.org/doi/10.1103/PhysRevB.110.165111}
}

@article{PhysRev.79.469,
  title = {Variational Principles for Scattering Processes. I},
  author = {Lippmann, B. A. and Schwinger, Julian},
  journal = {Phys. Rev.},
  volume = {79},
  issue = {3},
  pages = {469--480},
  numpages = {0},
  year = {1950},
  month = {Aug},
  publisher = {American Physical Society},
  doi = {10.1103/PhysRev.79.469},
  url = {https://link.aps.org/doi/10.1103/PhysRev.79.469}
}

@article{PhysRev.135.A130,
  title = {Electronic Structure of Alloys},
  author = {Beeby, J. L.},
  journal = {Phys. Rev.},
  volume = {135},
  issue = {1A},
  pages = {A130--A143},
  numpages = {0},
  year = {1964},
  month = {Jul},
  publisher = {American Physical Society},
  doi = {10.1103/PhysRev.135.A130},
  url = {https://link.aps.org/doi/10.1103/PhysRev.135.A130}
}

@article{doi:10.7566/JPSJ.83.014703,
  author = {Yanase ,Youichi},
  title = {Magneto-Electric Effect in Three-Dimensional Coupled Zigzag Chains},
  journal = {J. Phys. Soc. Jpn.},
  volume = {83},
  number = {1},
  pages = {014703},
  year = {2014},
  doi = {10.7566/JPSJ.83.014703},
  URL = {https://doi.org/10.7566/JPSJ.83.014703}
}

@article{Tursina:wm6046,
  author = "Tursina, Anna I. and Nesterenko, Sergei N. and Murashova, Elena V. and Chernyshev, Ilya V. and No{\"{e}}l, Henri and Seropegin, Yuri D.",
  title = "{CeRu${\sb 2}$Al${\sb {10}}$ with the YbFe${\sb 2}$Al${\sb {10}}$ structure type}",
  journal = "Acta Crystallographica Section E",
  year = "2005",
  volume = "61",
  number = "2",
  pages = "i12--i14",
  month = "Feb",
  doi = {10.1107/S1600536805000310},
  url = {https://doi.org/10.1107/S1600536805000310},
}

@article{doi:10.1143/JPSJ.80.073701,
  author = {Kato ,Harukazu and Kobayashi ,Riki and Takesaka ,Tomoaki and Nishioka ,Takashi and Matsumura ,Masahiro and Kaneko ,Koji and Metoki ,Naoto},
  title = {Magnetic Structure Determination of $\mathrm{Ce}{T}_2\mathrm{Al}_{10}$ (${T} = \mathrm{Ru}$ and $\mathrm{Os}$): Single Crystal Neutron Diffraction Studies},
  journal = {J. Phys. Soc. Jpn.},
  volume = {80},
  number = {7},
  pages = {073701},
  year = {2011},
  doi = {10.1143/JPSJ.80.073701},
  URL = {https://doi.org/10.1143/JPSJ.80.073701}
}

@article{Hayami_doi:10.7566/JPSJ.84.064717,
  author = {Satoru Hayami and  Hiroaki Kusunose and  Yukitoshi Motome},
  title = {{Spontaneous Multipole Ordering by Local Parity Mixing}},
  journal = {J. Phys. Soc. Jpn.},
  volume = {84},
  number = {6},
  pages = {064717},
  year = {2015},
  doi = {10.7566/JPSJ.84.064717},
}

@article{doi:10.7566/JPSCP.30.011189,
  author = {Shinozaki, Masahiro and Motoyama, Gaku and Mutou, Tetsuya and Nishigori, Shijo and Yamaguchi, Akira and Fujiwara, Kenji and Miyoshi, Kiyotaka and Sumiyama, Akihiko},
  title = {Study for Current-induced Magnetization in Ferrotoroidal Ordered State of {$\mathrm{Ce}_3\mathrm{TiBi}_5$}},
  year = {2020},
  volume = {30},
  pages = {011189},
  journal = {JPS Conf. Proc.},
  doi = {10.7566/JPSCP.30.011189},
  URL = {https://journals.jps.jp/doi/abs/10.7566/JPSCP.30.011189}
}

@article{doi:10.7566/JPSJ.89.033703,
  author = {Shinozaki ,Masahiro and Motoyama ,Gaku and Tsubouchi ,Masahiro and Sezaki ,Masumi and Gouchi ,Jun and Nishigori ,Shijo and Mutou ,Tetsuya and Yamaguchi ,Akira and Fujiwara ,Kenji and Miyoshi ,Kiyotaka and Uwatoko ,Yoshiya},
  title = {Magnetoelectric Effect in the Antiferromagnetic Ordered State of Ce3TiBi5 with Ce Zig-Zag Chains},
  journal = {J. Phys. Soc. Jpn.},
  volume = {89},
  number = {3},
  pages = {033703},
  year = {2020},
  doi = {10.7566/JPSJ.89.033703},
  URL = {https://doi.org/10.7566/JPSJ.89.033703}
}

@article{PhysRevResearch.3.023140,
  title = {High-temperature antiferromagnetism in $\mathrm{Yb}$ based heavy fermion systems proximate to a Kondo insulator},
  author = {Suzuki, Shintaro and Takubo, Kou and Kuga, Kentaro and Higemoto, Wataru and Ito, Takashi U. and Tomita, Takahiro and Shimura, Yasuyuki and Matsumoto, Yosuke and Bareille, C\'edric and Wadati, Hiroki and Shin, Shik and Nakatsuji, Satoru},
  journal = {Phys. Rev. Res.},
  volume = {3},
  issue = {2},
  pages = {023140},
  numpages = {12},
  year = {2021},
  month = {May},
  publisher = {American Physical Society},
  doi = {10.1103/PhysRevResearch.3.023140},
  url = {https://link.aps.org/doi/10.1103/PhysRevResearch.3.023140}
}

@article{sudo2025large,
  title = {Large Spontaneous Nonreciprocal Charge Transport in a Zero-Magnetization Antiferromagnet},
  author = {Sudo, Kenta and Yanagi, Yuki and Akaki, Mitsuru and Tanida, Hiroshi and Kimata, Motoi},
  journal = {Phys. Rev. Lett.},
  volume = {136},
  issue = {1},
  pages = {016503},
  numpages = {7},
  year = {2026},
  month = {Jan},
  publisher = {American Physical Society},
  doi = {10.1103/13pd-tlzp},
  url = {https://link.aps.org/doi/10.1103/13pd-tlzp}
}

@article{m96k-8ndj,
  title = {Revisiting magnetoelectric response in collinear antiferromagnetic zigzag chains: A downfolding approach beyond conventional low-energy models},
  author = {Kanda, Shuhei and Hayami, Satoru},
  journal = {Phys. Rev. B},
  volume = {114},
  issue = {16},
  pages = {165141},
  numpages = {16},
  year = {2026},
  month = {Sep},
  publisher = {American Physical Society},
  doi = {10.1103/m96k-8ndj},
  url = {https://link.aps.org/doi/10.1103/m96k-8ndj}
}

@article{Slater_PhysRev.94.1498,
  title = {Simplified LCAO Method for the Periodic Potential Problem},
  author = {Slater, J. C. and Koster, G. F.},
  journal = {Phys. Rev.},
  volume = {94},
  issue = {6},
  pages = {1498--1524},
  numpages = {0},
  year = {1954},
  month = {Jun},
  publisher = {American Physical Society},
  doi = {10.1103/PhysRev.94.1498},
  url = {http://link.aps.org/doi/10.1103/PhysRev.94.1498}
}

@article{PhysRevLett.102.146805,
  title = {Magnetoelectric Polarizability and Axion Electrodynamics in Crystalline Insulators},
  author = {Essin, Andrew M. and Moore, Joel E. and Vanderbilt, David},
  journal = {Phys. Rev. Lett.},
  volume = {102},
  issue = {14},
  pages = {146805},
  numpages = {4},
  year = {2009},
  month = {Apr},
  publisher = {American Physical Society},
  doi = {10.1103/PhysRevLett.102.146805},
  url = {https://link.aps.org/doi/10.1103/PhysRevLett.102.146805}
}

@article{PhysRevB.83.085108,
  title = {Chern-Simons orbital magnetoelectric coupling in generic insulators},
  author = {Coh, Sinisa and Vanderbilt, David and Malashevich, Andrei and Souza, Ivo},
  journal = {Phys. Rev. B},
  volume = {83},
  issue = {8},
  pages = {085108},
  numpages = {12},
  year = {2011},
  month = {Feb},
  publisher = {American Physical Society},
  doi = {10.1103/PhysRevB.83.085108},
  url = {https://link.aps.org/doi/10.1103/PhysRevB.83.085108}
}

@article{gmnv-cwvr,
  title = {Formulation of the orbital magnetic moment in multiorbital tight-binding models: Application to the inverse Faraday effect},
  author = {Tazuke, Kosuke and Morimoto, Takahiro and Kitamura, Sota},
  journal = {Phys. Rev. B},
  volume = {112},
  issue = {15},
  pages = {155134},
  numpages = {17},
  year = {2025},
  month = {Oct},
  publisher = {American Physical Society},
  doi = {10.1103/gmnv-cwvr},
  url = {https://link.aps.org/doi/10.1103/gmnv-cwvr}
}

@article{PhysRevB.111.125139,
  title = {Barycentric rational function approximation made simple: A fast analytic continuation method for Matsubara Green's functions},
  author = {Huang, Li and Yue, Changming},
  journal = {Phys. Rev. B},
  volume = {111},
  issue = {12},
  pages = {125139},
  numpages = {17},
  year = {2025},
  month = {Mar},
  publisher = {American Physical Society},
  doi = {10.1103/PhysRevB.111.125139},
  url = {https://link.aps.org/doi/10.1103/PhysRevB.111.125139}
}

@dataset{dataset_zenodo,
  author       = {Kanda, Shuhei and Hayami, Satoru},
  title        = {Dataset for: "Shape dependence of Edelstein and magnetoelectric effects in the V-shaped model"},
  year         = {2026},
  publisher    = {Zenodo},
  doi          = {10.5281/zenodo.23495185},
  url          = {https://doi.org/10.5281/zenodo.23495185},
}
% \bibliography{KunojiME_bib.bbl}

\end{document}